\documentclass{aa}  
\usepackage[table, dvipsnames]{xcolor}
\usepackage{setspace}
\usepackage{chngpage}
\usepackage{float}
\usepackage[english]{varioref}
\usepackage{booktabs}
\usepackage{hyperref}
\usepackage{siunitx}
\usepackage{siunitx}
\usepackage[normalem]{ulem}
\usepackage{caption}
\usepackage{graphicx}
\usepackage{txfonts}
\usepackage{lipsum}
\usepackage{subcaption}         
\usepackage{lscape}             
\usepackage{placeins}           

\begin{document}
\makeatletter
\let\linenumbers\relax
\let\runningpagewiselinenumbers\relax
\makeatother
   \title{Cepheid Metallicity in the Leavitt Law (C--MetaLL) survey: X.}

   \subtitle{First spectroscopic characterization of Galactic Type II Cepheids\\
   and calibration of the Period--Wesenheit--Metallicity relations}

%

   \author{T. Sicignano \inst{1,2,3,4}
        \and M. Romaniello \inst{1}
        \and A. Bhardwaj \inst{5}
        \and V. Ripepi \inst{3}
        \and M. Rejkuba \inst{1}
        \and M. Marconi\inst{3}
        \and E. Trentin\inst{3}
        \and G. Catanzaro \inst{6}
        \and G. De Somma \inst{3,4} 
        \and R. Molinaro \inst{3}   
        \and S. Das \inst{5}  
        \and I. Musella \inst{3} 
        \and L. Salmeri\inst{3,8} 
        \and P. Nishad \inst{5}  
        \and S. Kar \inst{5}
        \and S. Leccia \inst{3} 
        }

   \institute{European Southern Observatory, Karl-Schwarzschild-Strasse 2, 85748 Garching bei München, Germany 
   \and Scuola Superiore Meridionale, Largo San Marcellino10 I-80138 Napoli, Italy \and INAF-Osservatorio Astronomico di Capodimonte, Salita Moiariello 16, 80131, Napoli, Italy \and Istituto Nazionale di Fisica Nucleare, Sez. di Napoli, Monte S. Angelo, Via Cinthia Edificio 6 I-80126 Napoli, Italy \and Inter-University Center for Astronomy and Astrophysics (IUCAA), Post Bag 4, Ganeshkhind, Pune 411 007, India \and INAF-Osservatorio Astrofisico di Catania, via Santa Sofia 78, I-95125 Catania, Italy \and Università degli studi di Napoli Federico II, Dipartimento di Fisica "Ettore Pancini", via Cinthia 21, Edificio 6, Napoli, Italy
}

   \date{Received September 30, 20XX}

 
  \abstract
   {Type II Cepheids (T2Cs) are Population II pulsating stars that can be used as primary distance indicators thanks to their period--luminosity relation. Together with RR Lyrae stars and the tip of the red giant branch, they can provide an alternative route to the Hubble constant, independently evaluating possible systematic effects in its determination. 
    }
   {We aim to build the largest homogeneous spectroscopic sample of Galactic T2Cs with iron and oxygen abundance measurements to investigate the impact of chemical composition on their period--luminosity relations.
   } 
   { We observed 80 Galactic T2Cs with the high-resolution spectrograph UVES at the ESO Very Large Telescope. Starting from the spectra for a sample of 74 T2Cs we derived stellar parameters, iron and oxygen abundances through a self-consistent spectroscopic analysis based on equivalent width of absorption lines.
   This sample almost doubles the number of T2Cs with spectroscopic metallicity determinations available in the literature.
   }
   { The resulting abundance measurements were used to test theoretical predictions for the instability strip and, together with $Gaia$ parallaxes, to derive Period--Wesenheit--metallicity (PWZ) relations. We find, in agreement with model predictions, that the metallicity term ($\gamma$) is small and statistically consistent with zero in all investigated PWZ relations, ranging between $\gamma \simeq -~0.02$ and $+0.17$ mag dex$^{-1}$. We also estimated distances to 22 Galactic Globular Clusters hosting T2Cs. The derived distances agree, within the uncertainties, with independent literature estimates.
   }
   {Our results indicate that metallicity has at most a minor impact on the pulsation properties and distance scale of T2Cs, confirming recent theoretical predictions and supporting the use of these variables as robust Population II distance indicators.}


   \maketitle

\section{Introduction} \label{intro}
In the last century, $H_0$ measurements based on both early and late universe observations have become more accurate and precise, resulting in the so-called "Hubble tension" \citep[see e.g.][]{2021APh...13102605D,2026h0dn}. The tension arises because of a 5-7 $\sigma$ difference in the $H_0$ values coming from two most accurate direct and indirect determinations \citep[][]{cosmoverse2025}: the first one is based on Cosmic Microwave Background Radiation observations analysed together with the Lambda Cold Dark Matter model of the Universe \citep[][]{Planck2020}, the other is a direct measurement of Type Ia Supernovae in the Hubble flow calibrated with Classical Cepheids (CCs) distances in the Local Universe \citep[][]{Riess2022}. Evaluating the significance of this tension warrants the use of alternate distance tracers and independent approaches often involving multiple probes to control systematic uncertainties \citep[e.g.,][]{2026h0dn}. Population II distance indicators, such as RR Lyrae (RRL) stars and Type II Cepheids (T2Cs), together with the tip of the Red Giant Branch (TRGB), offer the possibility of providing an alternative route to $H_0$ \citep[][]{Beaton2016,bhardwaj2022rr,2026A&A...708L...8C}.

T2Cs, being older and less massive than CCs, populate various environments: the Galactic Bulge \citep[see e.g.][]{soszynski11,soszynski17,matsunaga13,braga19}, the Galactic Globular Clusters \citep[GGCs, see][]{matsunaga2006,ngeow22}, the Galactic halo \citep[][]{Ripepi2023}, and the Magellanic Clouds \citep[see e.g.][hereinafter S24]{ripepi2015,soszynski2018ogle,sicignano24}. 
T2Cs are brighter by more than 0.2 dex than RRL and, according to their pulsation period, are classified into three sub-groups \citep[see e.g.][]{soszynski08,bono20,bono2024}: BL Herculis (BLHer) stars with periods longer than RRLs (1 day) and shorter than four days; W Virginis (WVir) stars with periods between four and twenty days, RV Tauri stars with periods longer than twenty days. However, these period boundaries are only approximate and vary with the metallicity of their host stellar system. Moreover, a subsample of WVir stars showing peculiar light curves and brighter magnitudes is suspected to have a binary origin \citep[][]{soszynski08}.

T2Cs can be used as distance indicators because they follow very tight period--luminosity, period--Wesenheit\footnote{The Wesenheit magnitudes are constructed to be reddening free by definition \citep[][]{madore1982period}}, and period--luminosity--colour relationships \citep[PL, PW and PLC, respectively; see e.g.][S24]{feast2008luminosities,matsunaga2011,ripepi2015,Bhardwaj2017_LMC,bhardwaj2020, cruzreyes2025, lengen2025}. Recently, in S24, new empirical calibrations of T2Cs PL and PW relations were used to obtain distances to 22 GGCs, supporting the combined use of these pulsating stars and the TRGB for cosmic distance scale studies \citep[see also][and references therein]{Baade58,Bono16,Bhardwaj2021}. To assess T2Cs as primary distance indicators, it is important to establish the metallicity dependence of their PL and PW relations, as extensively argued by \citet{das2026}. Although the theoretical results predict weak or negligible metallicity dependence, in particular for the BLHer subclass \citep[][]{marconi07,dicriscienzo07,Deka2024,marconi2025}, observations are not statistically sufficient to draw firm conclusions \citep[][]{ngeow22,Wielgorski2022}. The increasingly accurate and precise $Gaia$ parallaxes \citep[][]{GaiaPrusti2016} offer the best opportunity to quantify metallicity effects on the Period-Wesenheit-Metallicity (PWZ) relations for Galactic field T2Cs. The main limiting factors for a precise calibration of PWZ relations are now the lack of homogeneous multiband time-series photometry and spectroscopic metallicities ([Fe/H]) for a statistically significant sample.

The Cepheids-Metallicity in the Leavitt Law survey \citep[C-MetaLL,][]{Ripepi2021} was conceived to quantify the metallicity effect on the Leavitt law \citep[][]{leavitt1912periods}. The main case study has been the CCs, enriching their spectroscopic sample with the most metal-poor galactic CCs \citep[][]{Ripepi2021,trentin2023,trentin2024b,Trentin2026}. Complementary multiband light curves of CCs have also been obatained as part of the C-MetaLL survey \citep[][]{bhardwaj2024}. In this paper on alternative primary calibrators in the context of the C-MetaLL project, we present the first high-resolution spectra for a sizable sample of T2Cs and their empirical PWZ relations in a variety of bands. Follow-up studies will present homogeneous multi-band time-series photometry and further expand the sample of spectroscopic observations of Galactic T2Cs.


The paper is structured as follows: in Sect.~\ref{description} we present the photometric and spectroscopic data, together with their relative reduction to calibrated one-dimensional spectra. Section~\ref{abund analysis} focuses on the derivation of the stellar parameters for the target stars, the derived abundances and the comparison with theoretical predictions.  
In Sect.~\ref{metdep} we calculate, calibrate and discuss the T2C PWZ relations, also in comparison with theoretical predictions and observations. Section~\ref{applications} presents the GGCs distance moduli based on our new T2Cs PWZ. Finally, conclusions are drawn in Sect.~\ref{conclus}.

\section{Description of the data sample} \label{description}
\subsection{Selection of targets}
The $Gaia$ Data Release 3 \citep[DR3,][]{GaiaVallenari2023} provides medium resolution Gaia-RVS spectroscopic metallicities for 166 Galactic T2Cs as discussed in detail by \citet[][]{2024IAUbhardwaj}, but the compiled sample of high-resolution spectra of T2Cs from the literature \citep[e.g.][]{maas2007,luck2011, Kovtyukh2018} is not homogeneous and includes only 19 short-period T2Cs. Furthermore, a discrepancy of up to 0.3 dex is present between Gaia-RVS and literature based on high-resolution spectra. This suggests that the Gaia-RVS metallicities are not reliable for these metal-poor variable stars. Among the aforementioned 166 T2Cs, there are 84 BLHer and WVir stars, all of which have accurate and precise (errors < 10\%) parallaxes. RV Tau stars were not considered in this study, given their large parallax uncertainties and the fact that they follow a different period-luminosity relation than shorter-period Type II Cepheids \citep[][]{ripepi2015, sicignano24}.  

\begin{figure*}
\centering
    \includegraphics[width=15cm]{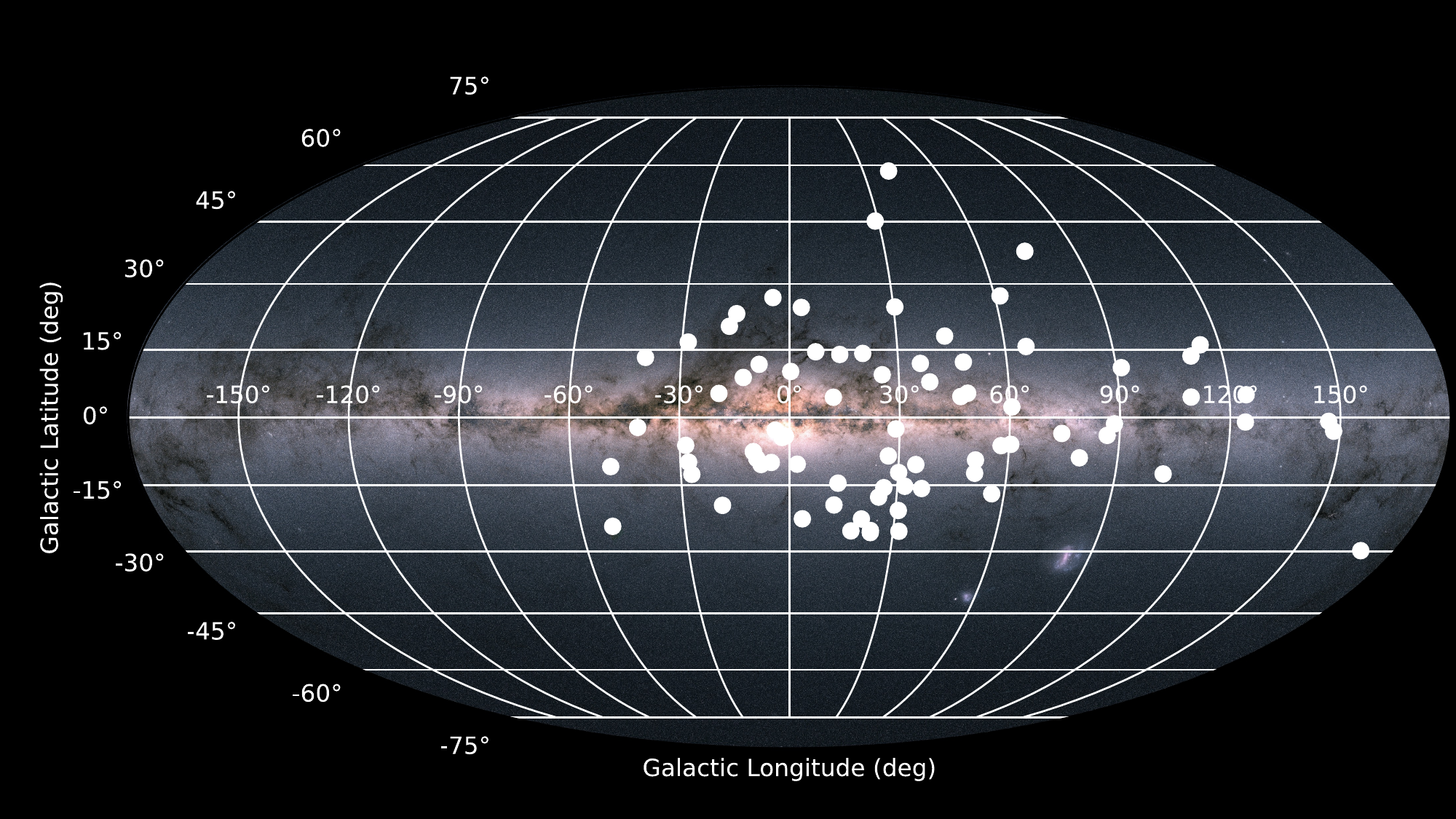}
    \caption{\label{map}  Distribution of the 80 T2Cs in the Milky Way. Credits for the background image: ESA/Gaia/DPAC, CC BY-SA 3.0 IGO.  } 
        \end{figure*}

In the context of the C-MetaLL project, our goal is to collect high-resolution spectra for these 84 T2Cs and then to extend the sample, once $Gaia$ Data Release 4 is available, providing additional targets with accurate and precise parallaxes. 
The dataset analysed in this first work comprises 80 targets, shown in Figure~\ref{map}, for which we obtained high-resolution spectra (see Sect.~\ref{sect:spectroscopic observations}). However, during the pre-analysis of the observed spectra, the sample was cleaned up excluding suspected binaries. Therefore, the final sample presented in this work is made of 74 T2Cs. 
Table~\ref{astrotable} lists the identification and the main properties of the targets, including the periods and epochs of maximum light in the $Gaia$ $G$ band \citep[][]{Ripepi2023} and the DR3 parallaxes with the bias corrections provided by \citet{Lindegren2021}. 

\begin{table*}
\begin{center}
\caption{Identification and main parameters of the T2Cs targets.  
\label{astrotable}} 
\footnotesize\setlength{\tabcolsep}{3pt}
 
\begin{tabular}{cccccccccc}
\hline
\noalign{\smallskip}   
ID $Gaia$ & ID SIMBAD & RA & DEC & Class & P & Epoch & $\varpi$ & $\sigma_{\varpi}$ & $\varpi_{bias}$ \\
& & deg & deg & & days & days & mas & mas & mas \\
(1) & (2) & (3) & (4) & (5) & (6) & (7) & (8) & (9) & (10) \\
\noalign{\smallskip}
\hline
\noalign{\smallskip}

6109579880366892416 & ASAS\_J140159-4256.3 & 210.494 & $-$42.938 & BLHer & 1.153 & 56896.6044 & 0.358 & 0.017 & $-$0.028 \\
4344611414225212288 & ASAS\_J162326-0941.0 & 245.860 & $-$9.683 & BLHer & 1.378 & 56906.8199 & 1.410 & 0.017 & $-$0.034 \\
4076363528037529600 & V2022\_Sgr & 280.162 & $-$25.381 & BLHer & 1.533 & 56910.8606 & 0.313 & 0.020 & $-$0.044 \\
5796440725425239936 & UCAC2 895326 & 214.651 & $-$74.201 & WVir & 5.328 & 56887.0492 & 0.384 & 0.014 & $-$0.019 \\
5620149154199221632 & OGLE-GD-T2CEP-0009 & 112.850 & $-$20.775 & WVir & 12.551 & 56927.0093 & 0.137 & 0.012 & $-$0.038 \\

\noalign{\smallskip}
\hline
\noalign{\smallskip}
\end{tabular}
\tablefoot{Columns: $(1)$ $Gaia$ source id; $(2)$ Identification from Simbad;
$(3)$-$(4)$ RA and Dec;
$(5)$ Class: BLHer = BL Herculis; WVir = W Virginis;
$(6)$ Period from $Gaia$ DR3;
$(7)$ Epoch of maximum light in the $Gaia$ $G$ band;
$(8)$-$(9)$-$(10)$ parallax, its uncertainty, and parallax bias from $Gaia$ according with \citet[][]{Lindegren2021}; The Z5 bias function was adopted for all targets except 5828480459918679936, 5845741383716281216, 5855676944429471872, 6434564460631076864, 6643297500393311616, and 6704322216802230784, for which the Z6 function was used.
A portion is shown here for guidance regarding its form and content. The machine-readable version of the full table will be published at the \href{https://cds.u-strasbg.fr/}{Centre de Données astronomiques de Strasbourg (CDS)}.}
\end{center}
\end{table*}

\subsection{Photometry}
$Gaia$ photometry is available for the complete sample \citep[][]{Ripepi2023}, while the Optical Gravitation Lensing Experiment \citep[OGLE][]{udalski2018} optical V and I photometry is available for 10 and 16 objects, respectively\citep[][]{soszynski17,soszynski2018ogle,soszynski2020}. The photometric transformations proposed by \citet{pancino2022gaia} were used to complement the $V\ I$ photometry starting from the $Gaia$ bands; in particular, according to \citet{trentin2024a}, the calculated $I$ photometry resulted showed a bias of 0.035 mag and the I band magnitudes were increased accordingly to be comparable with OGLE $I$ band. 


The 2MASS single epoch photometry \citep[][]{skrutskie2006} for the $J$ and $K_s$ bands has been transformed in mean-magnitude following the template technique introduced by \citet{Ripepi2021} and using templates of T2C light curves available in S24. 
The multi-band photometry obtained as described above is reported in Table~\ref{photometry}.

\begin{table*}
\begin{adjustwidth}{-0.8cm}{}
\begin{center}

\caption{Photometry of the targets. 
\label{photometry}} 
\small
\footnotesize\setlength{\tabcolsep}{3pt}
\begin{tabular}{rrrrlrrrrrrrrrrlrrrr}
\hline
\noalign{\smallskip}   
ID$Gaia$ &  RA & DEC & P & $\langle G \rangle$ & $\sigma_{\langle G \rangle}$& $\langle G_{BP} \rangle$ & $\sigma_{\langle G_{BP} \rangle}$ & $\langle G_{RP} \rangle$ & $\sigma_{\langle G_{RP} \rangle}$  & $\langle V \rangle$ & $\sigma_{\langle V \rangle}$ & $\langle I \rangle$ & $\sigma_{\langle I \rangle}$ & VI & $\langle J \rangle$ & $\sigma_{\langle J \rangle}$&  $\langle K \rangle$ & $\sigma_{\langle K \rangle}$ & Qfl \\
 & deg & deg &days & mag &mag & mag & mag & mag &mag & mag & mag & mag&mag & & mag & mag & mag &mag &  \\
(1)  & (2)  & (3) & (4) & (5) & (6) & (7) &(8) & (9) & (10) & (11) & (12)  & (13) & (14) & (15) &(16)  & (17) &(18) & (19) & (20) \\        
 \noalign{\smallskip}
  \hline  
  \noalign{\smallskip}  
892416 & 210.4943 & $-$42.9384 & 1.153 & 12.333 & 0.002 & 12.617 & 0.003 & 11.879 & 0.015& 12.474 & 0.005 & 11.85 & 0.07 & P,P & 11.55 & 0.05 & 11.25 & 0.04 & AAA \\  
212288 & 245.8600 & $-$9.6832 & 1.378 & 9.587 & 0.002 & 9.998 & 0.001 & 8.999 & 0.000  & 9.804 & 0.002 & 8.95 & 0.04 & P,P & 8.31 & 0.05 & 7.90 & 0.04 & AAA \\
529600 & 280.1624 & $-$25.3806 & 1.533 & 13.262 & 0.022 & 13.703 & 0.026 & 12.616 & 0.015  & 13.511 & 0.025 & 12.52 & 0.02 & P,O & 11.96 & 0.15 & 11.46 & 0.08 & AAA \\
239936 & 214.6507& $-$74.2010 & 5.328 & 11.430 & 0.002 & 11.932 & 0.005 & 10.751 & 0.004  & 11.716 & 0.004 & 10.69 & 0.04 & P,P & 9.97 & 0.04 & 9.38 & 0.03 & AAA \\
221632 & 112.8501 & $-$20.7748 & 12.551 & 13.293 & 0.003 & 14.114 & 0.004 & 12.393 & 0.006 & 13.94 & 0.02 & 12.33 & 0.02 & O,O  & 11.31 & 0.09 & 10.51 & 0.07 & AAA\\
 \noalign{\smallskip}
  \hline  
  \noalign{\smallskip}  
\end{tabular}

\tablefoot{Columns: $(1)$ Identification from $Gaia$ restricted to the last six digits (in the full table the complete ID will be present) ; $(2)$-$(3)$ RA and Dec; $(4)$ Period from $Gaia$; $(5)$-$(6)$, $(7)$-$(8)$, $(9)$-$(10)$ magnitudes and relative uncertainties in $G$, $G_{BP}$, and $G_{RP}$ bands from $Gaia$; $(11)$-$(12)$, $(13)$-$(14)$ magnitudes and relative uncertainties in $V$ and $I$ bands, column $(15)$ specifies V (first) and I (second) photometry sources: P refers to transformations from \citet{pancino2022gaia}, O refers to the \href{https://ogledb.astrouw.edu.pl/~ogle/OCVS/t2cep_query.php}{OGLE catalogue}; $(16)$-$(17)$, $(18)$-$(19)$ template-averaged magnitudes and relative uncertainties in $J$ and $K$ bands starting from 2MASS single-epoch photometry; $(20)$ quality flag of the 2MASS single-epoch photometry. The machine-readable version of the full table will be published at the \href{https://cds.u-strasbg.fr/}{CDS}.}
\end{center}  
\end{adjustwidth}{}{}
\end{table*}

\subsection{Spectroscopy}
\label{sect:spectroscopic observations}
The spectra analysed in this work have been obtained in the context of two ESO (European Southern Observatory) proposals, namely P115.28AS and P116.291H (P.I. Bhardwaj).
The observations were taken between April 2025 and March 2026 in service mode.

We used the Ultraviolet and Visual Echelle Spectrograph \citep[UVES,][]{dekker2000} instrument, attached to the Unit Telescope 2 of Very Large Telescope (VLT) on Cerro Paranal (Chile). We adopted the red arm equipped with the grism CD$\#$3, covering the wavelength interval 4760-6840\,\AA, and the central wavelength at 5800\,\AA. We employed the 1 arcsec slit, which provides a resolution of R$\sim$47000. Observing constraints and exposure times were adapted to each star in order to deliver data of uniform quality (signal-to-noise ratios higher than 40 at 587 nm). No constraints were set on the pulsational phase at the time of observation, which is therefore random. 
The data reduction was carried out using the standard UVES pipeline version 6.6.1 \citep{uves_pipe} through the ESO Data Processing System \citep[EDPS][]{edps}.  Spectra of the same object in the same night were stacked with the \textit{esotk\_spectrum1d\_combine} recipe included in the ESO toolkit pipeline esotk-1.0.0 \citep{esotk}. This procedure provides two fully calibrated 1D spectra for each exposure, one for each wavelength interval covered by the instrument. Three stars (namely 6045677196574103296, 6638377559519795456, and 6706838857179992704) were observed on two different nights because the initial observations did not fully meet our observational constraints. After visually inspecting the signal-to-noise ratios of the spectra, we decided to retain these observations and use them for a consistency check of our spectral analysis.


\begin{figure}
\centering
    \includegraphics[width=9cm]{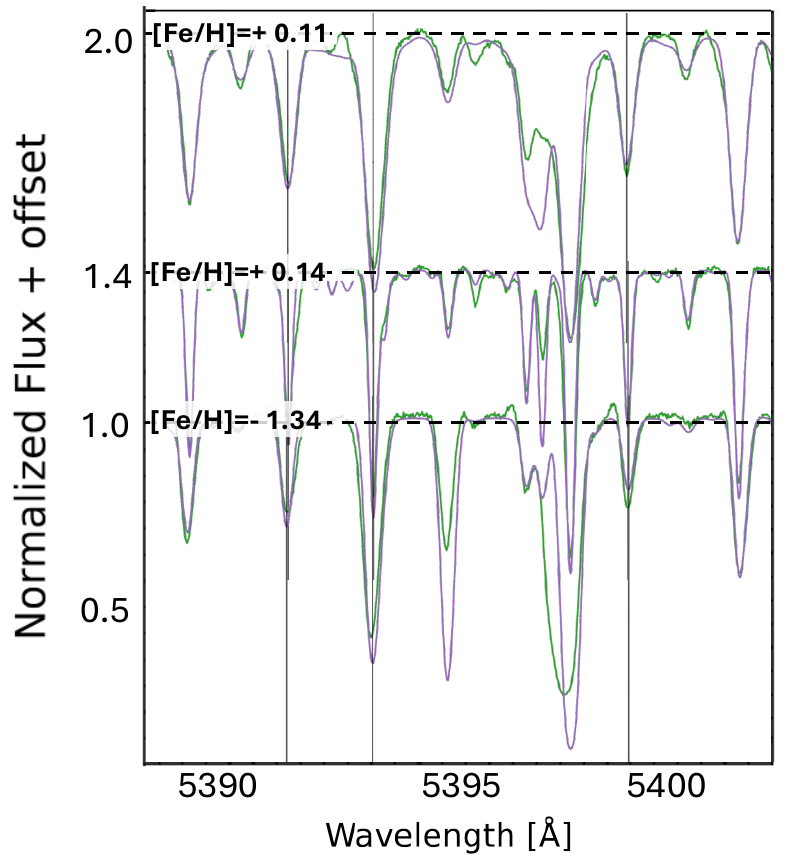}
    \caption{\label{spectra} Example of an observed high-resolution spectrum (green) and a synthetic spectrum (purple) computed using the stellar parameters derived with GALA from EWs, over the wavelength range $\lambda \lambda$ 5390–5400\AA. Black vertical lines mark the positions of FeI lines used by GALA.}          \end{figure}

\section{Abundance analysis} \label{abund analysis}
The chemical abundances were derived through the equivalent width (EW) method.
The EWs of Fe I and Fe II lines, adopted from \citet[][]{Romaniello2022A&A...658A..29R} and compiled specifically for the parameter space of Cepheids, were measured with the DAOSPEC code \citep[][]{2008DAOSPEC} thanks to the 4DAO software \citep[][]{2013Mucciarelli}. Our instrumental setup includes only very few oxygen lines making the oxygen abundances more uncertain than the iron ones. Specifically, we used the two forbidden [OI] lines at 6300.30 \AA\ (see Figure \ref{oxygen}) and 6363.77 \AA\ to measure the oxygen abundances for our stars.

Stellar parameters, namely the effective temperature ($T_{\rm eff}$), surface gravity ($\log\,g$), and microturbulent velocity ($\xi$),  are needed to create a stellar atmosphere model from which measured EWs are compared with the observed EWs. The set of final stellar parameters comes from an iterative procedure that converges when the following conditions are met:

\begin{itemize}
    \item excitation equilibrium: $T_{\rm eff}$ is derived by imposing that there should be no residual correlation between the iron abundance and the excitation potential $\chi$ of the neutral iron lines. 
    \item ionisation equilibrium: $\log\,g$ is derived by requiring that, for a given species, the abundance is the same (within the uncertainties) from lines of two different ionisation states (in our case, neutral and singly ionised iron lines).
    \item no residual correlation between the iron abundance and the line EW as a measure of the line strength to set $\xi$.
\end{itemize}
The initial values of the parameters were set for all stars as follows: $T_{\rm eff}$ = 5000 K, $\log\,g$ = 1 cm/s$^2$, [Fe/H] = -0.1 dex and $\xi$ = 2 km/s.

In our analysis, we adopted the GALA implementation of the equivalent-width (EW) method, which has been extensively tested on UVES data \citep[][]{2013gala}. As input physics, we used the classical grid of ATLAS9 Local Thermodynamic Equilibrium (LTE) stellar atmosphere models \citep[][]{2003castelli}. Although these models cannot fully describe the complex and dynamic atmospheres of T2Cs, they provide a suitable framework for our analysis of iron and oxygen lines (see Figures \ref{spectra}, \ref{spectra2}, and \ref{spectra3}). This is further supported by the fact that the three stars observed at two different pulsation phases yield consistent chemical abundances (see Table \ref{diffphas}), suggesting that our approach provides robust abundance estimates despite the limitations of the LTE assumption. Notably, one of these stars was observed both during a relatively quiescent atmospheric phase and at a phase close to the onset of atmospheric shocks. Nevertheless, the agreement between the derived abundances does not rule out possible phase-dependent effects of atmospheric dynamics, such as shocks, on other spectral features. A dedicated analysis based on multi-epoch spectroscopy will be needed to investigate these effects in greater detail.

The final set of lines used in the analysis varies from one spectrum to another, as individual lines are rejected according to several quality criteria, including signal-to-noise ratio and line depth. Nevertheless, a sufficient number of lines is retained in each case to obtain reliable estimates of the stellar parameters and iron abundances (see Figure \ref{nlines}). The final iron abundance is the mean of the iron abundances from the individual lines in the convergence iteration. The same applies for oxygen abundance. We were able to measure the oxygen abundances for a subset of 61 stars. For 13 stars the EW of the lines could not be measured or convergence was not reached in deriving the abundance.

Once convergence is reached, the uncertainties in the elemental abundances are computed by accounting for the covariance among the stellar parameters, following the prescription of \citet[][]{2004cayrel} (see Appendix \ref{diagnostics} for further details).

\begin{figure}
\centering
\hspace*{-0.3cm}
    \includegraphics[width=10cm,clip,trim=2.1cm 0cm 1.2cm 2.6cm]{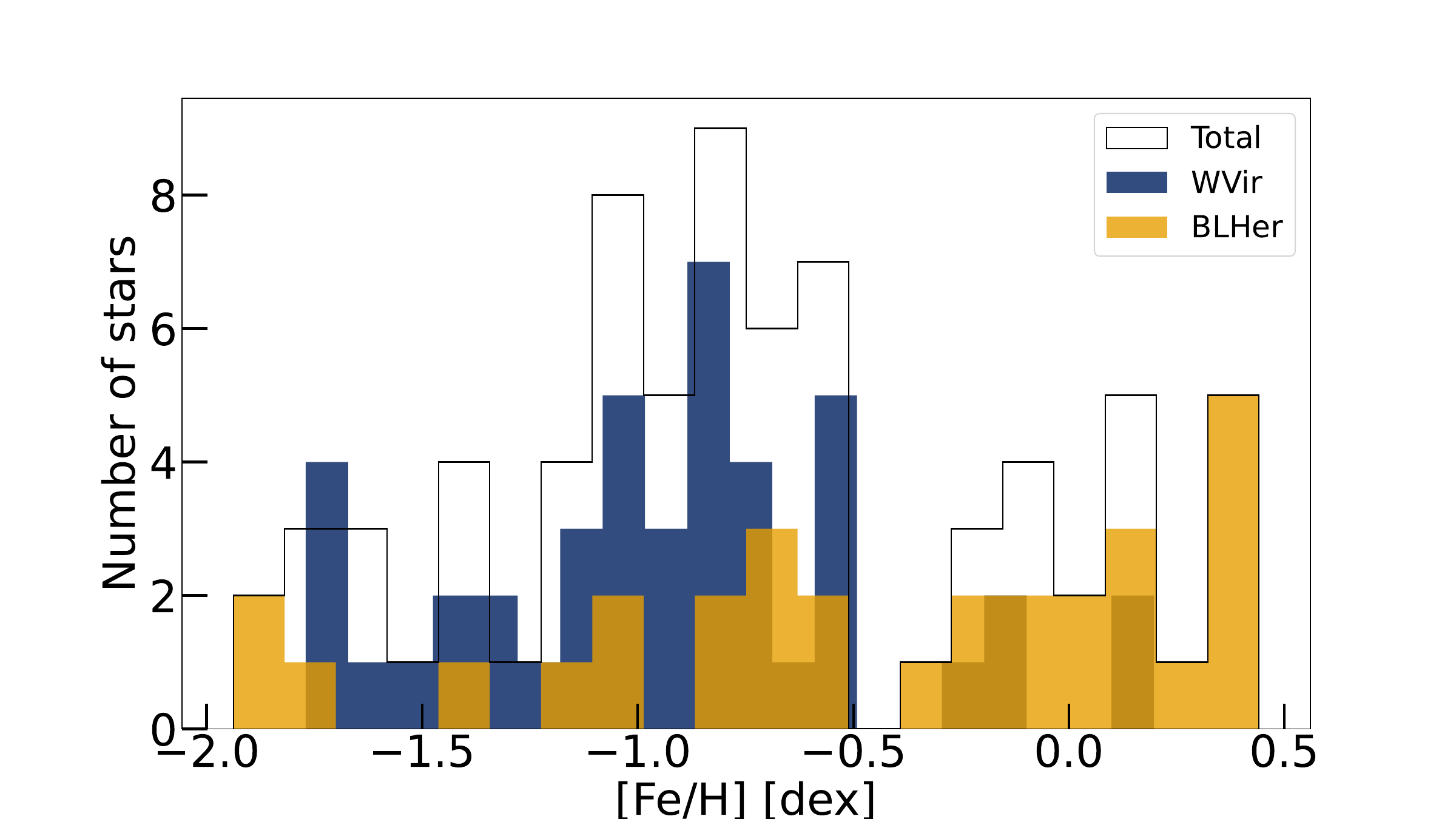}
    \caption{\label{hist} Metallicity distribution of the sample: yellow, blue and open histograms represent the BLHer, WVir and BLHer+WVir samples, respectively.
   } 
        \end{figure}

The estimated atmospheric parameters are summarised in Table~\ref{par}. 
Figure~\ref{hist} shows the metallicity distribution of our targets. Oxygen and iron abundances are reported together with the atmospheric parameters in Table~\ref{par}. 

\begin{table*}
\begin{center}
\caption{Spectroscopic parameters and metal abundances of the targets. 
\label{par}} 
\footnotesize\setlength{\tabcolsep}{3pt}
\begin{tabular}{rrrrrrrrr}
\hline
\noalign{\smallskip}   
ID $Gaia$ & $\Phi_{\rm Obs}$ & $MJD_{\rm Obs}$ & SNR & $T_{\rm eff}$ & $\xi$ & $\log\,g$  & [O/H] & [Fe/H] \\
 & & d & @5870 \AA & K & km/s & dex & dex & dex  \\
(1)  & (2)  & (3) & (4) & (5) & (6) & (7) &(8) & (9)  \\        
 \noalign{\smallskip}
  \hline  
  \noalign{\smallskip}  
6109579880366892416 & 0.62 & 61064.29297 & 138 & 6130$\pm$ 101 & 3.0 $\pm$ 0.1& 2.2 $\pm$ 0.1 & 0.68 $\pm$ 0.09 & 0.44 $\pm$ 0.07 \\
4344611414225212288 & 0.96 & 61104.37266 & 67 & 6230 $\pm$ 149 & 2.5 $\pm$ 0.1 & 1.8 $\pm$ 0.1  &    & 0.173 $\pm$ 0.09 \\
4076363528037529600 & 0.27 & 60954.07964 & 102 & 5560 $\pm$ 89 & 2.3 $\pm$ 0.08 & 1.1 $\pm$ 0.1 &  $-$0.21 $\pm$ 0.10  & $-$0.72 $\pm$ 0.05 \\
5796440725425239936 & 0.56 & 61029.33267 & 55 & 5240 $\pm$ 33 & 3.3 $\pm$ 0.1 & 1.2 $\pm$ 0.1 &  0.44 $\pm$ 0.05  & $-$0.18 $\pm$ 0.02 \\
5620149154199221632 & 0.38 & 60960.31512 & 91 & 4720 $\pm$ 62 & 2.9 $\pm$ 0.1 & 0.0 $\pm$ 0.07 &  $-$0.69 $\pm$ 0.07  & $-$0.80 $\pm$ 0.04 \\

 \noalign{\smallskip}
  \hline  
  \noalign{\smallskip}  
\end{tabular}

\tablefoot{Columns: $(1)$ Identification from $Gaia$; $(2)$ corresponding phase of the observation; $(3)$ Modified Julian day of the observation; $(4)$ spectral signal to noise ratio (SNR) at 5870 \AA; $(5)$ effective temperature; $(6)$ microturbolence velocity; $(7)$ $\log\,g$ as a proxy of surface gravity; $(8)$ Oxygen  logarithmic abundance with respect to hydrogen, in solar units: $[Fe/H] \equiv A(O)_{star} - A(O)_{\odot}$, where $A(O) = \log(N_{O}/N_H)$ + 12 and we adopt $A(=)_{\odot} \equiv$ 8.69 from \citet[][]{2009asplund}; $(9)$ Iron logarithmic abundance with respect to hydrogen, in solar units: $[Fe/H] \equiv A(Fe)_{star} - A(Fe)_{\odot}$, where $A(Fe) = \log(N_{Fe}/N_H)$ + 12 and we adopt $A(Fe)_{\odot} \equiv$ 7.50 from \citet[][]{2009asplund}. A portion is shown here for guidance regarding its form and content. The machine-readable version of the full table will be published at the \href{https://cds.u-strasbg.fr/}{CDS}.}
\end{center}
\end{table*}

Our aim is to calibrate the T2C PWZ relations and investigate their dependence on metallicity; therefore, in the following sections we will mainly consider the iron abundance as a proxy for the overall metallicity. Oxygen abundances will become relevant when measuring distances to distant galaxies, where oxygen is commonly used as a proxy for the overall metallicity. 

\subsection{Stellar versus pulsational parameters}
\begin{figure}
\centering
    \includegraphics[width=9cm,clip,trim=0cm 0.09cm 0cm 0cm]{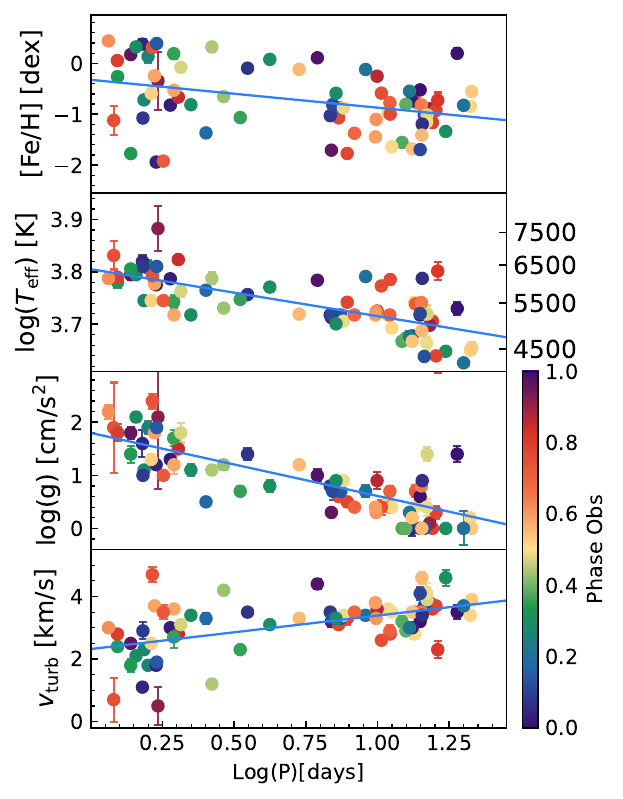}
    \caption{\label{stellvspuls} Stellar parameters (from top to bottom: Iron abundance, effective temperature, surface gravity, microtubolence velocity) of the targets as measured in our analysis vs the pulsational periods of the stars. The points are colour-coded according to the phase along the pulsational cycle within which the stars were observed, with the scale represented by the bar on the far right. Error-bars are plotted for all the points, but are often smaller than the point size. } 
        \end{figure}

Figure~\ref{stellvspuls} shows the stellar parameters as a function of pulsation period, color-coded according to the observed phase. The trends observed for T2Cs are similar to those found for CCs and presented in Fig.~6 of \citet[][]{Romaniello2022A&A...658A..29R}.

The iron abundance is an intrinsic property of the star independent of the pulsation mechanism, and this is reflected in its negligible correlation with the period. 
On the other hand, correlations are found between the period and the other stellar parameters. In particular, the $T_{\rm eff}$ steadily decreases as the period increases, as expected from the period-mass-luminosity-effective temperature relation which holds for T2C pulsators \citep[see e.g.][]{marconi2025}. 
While $\log\,g$ decreases as the period increases, because longer period pulsators are more luminous and have a more rarefied atmosphere. 
As expected \citep[Chap. 17][]{Gray_2021}, micro-turbulent velocities increase toward longer periods, as a direct consequence of the structural expansion and lower density in the atmospheres of more evolved stars. This systematic trend is well-documented in the literature for pulsating variables \citep[e.g.,][]{kovtyukh2008}. For the T2Cs in our sample, this correlation becomes particularly steep when transitioning from the more compact BLHer stars to the more expandend longer-period WVir subclass (moving toward lower $\log\,g$ values). Indeed, WVir pulsating stars are already known in the literature for their complex hydrodynamics and bimodal velocity of atmospheres \citep[][]{kovtyukh2011}. 
We note that the scatter in the stellar parameters at fixed period (see last three panels of Fig.~\ref{stellvspuls}) may be partially explained by the fact that the spectra have been taken at random phases along the pulsational cycle.

\subsection{Comparison with literature}

Ten of our targets overlap with previous studies that derived iron abundances from high-resolution spectra \citep{maas2007,luck2011,Kovtyukh2018}. The comparison is presented in Table~\ref{tab:comp_fe}. The metallicity values reported by \citet{maas2007}, who adopted a method for determining stellar parameters similar to ours, differ from our estimates by less than 0.3 dex. More pronounced differences are found when comparing our results with those of \citet{luck2011} and \citet{Kovtyukh2018}, who both employed different spectroscopic analysis techniques. Given the small number of stars in common between our sample and those of previous studies, we cannot determine whether these discrepancies reflect systematic offsets between the different analysis methods. Future studies based on larger samples will allow a more robust assessment of potential systematic differences among spectroscopic analysis techniques.

\begin{table}[!ht]
\begin{adjustwidth}{-0.6cm}{}
\centering
\small
\renewcommand{\arraystretch}{1.3}
\caption{Comparison of [Fe/H] with literature estimates.}
\label{tab:comp_fe}
\begin{tabular}{lrrrr}
\hline
Star &
$[Fe/H]_{S26}$ &
$\Delta_{S26-M07}$ &
$\Delta_{S26-K18}$ &
$\Delta_{S26-L11}$ \\
& dex & dex & dex & dex \\
\hline
SW Tau       & $+0.14 \pm 0.13$ & $-0.01 \pm 0.18$ &                    &                    \\
TX Del       & $+0.11 \pm 0.07$ & $+0.06 \pm 0.11$ &                    &                    \\
V526 Aql     & $+0.08 \pm 0.08$ &                    &                    & $-0.42 \pm 0.13$ \\
V553 Cen     & $-0.08 \pm 0.09$ &                    & $-0.16 \pm 0.13$ &                    \\
$\kappa$ Pav & $-0.12 \pm 0.08$ &                    & $-0.29 \pm 0.13$ &                    \\
VY Pyx       & $-0.26 \pm 0.10$ & $+0.09 \pm 0.16$ &                    &                    \\
AL Vir       & $-0.59 \pm 0.05$ & $-0.14 \pm 0.10$ & $-0.29 \pm 0.11$ &                    \\
BC Aql       & $-0.65 \pm 0.04$ &                    &                    & $-0.37 \pm 0.11$ \\
CO Pup       & $-0.92 \pm 0.05$ & $-0.27 \pm 0.15$ &                    &                    \\
W Vir        & $-1.34 \pm 0.03$ & $-0.29 \pm 0.13$ & $-0.76\pm 0.10$ &                    \\
\bottomrule
\end{tabular}
\tablefoot{ The differences are defined as $\Delta_{S26-X}=[Fe/H]_{S26}-[Fe/H]_X$; their uncertainties are propagated in quadrature. M07 refers to \citet{maas2007} (the adopted solar iron abundance is revised from $A(Fe)_{\odot}=7.45$ to $7.50$ ), K18 to \citet{Kovtyukh2018} (the adopted solar iron abundance is revised from $A(Fe)_{\odot}=7.57$ to $7.50$ ), L11 to \citet{luck2011}, and S26 to this work.}
\end{adjustwidth}
\end{table}


\subsection{Comparison with theory}
\begin{figure}
\centering
    \includegraphics[width=9cm,clip]{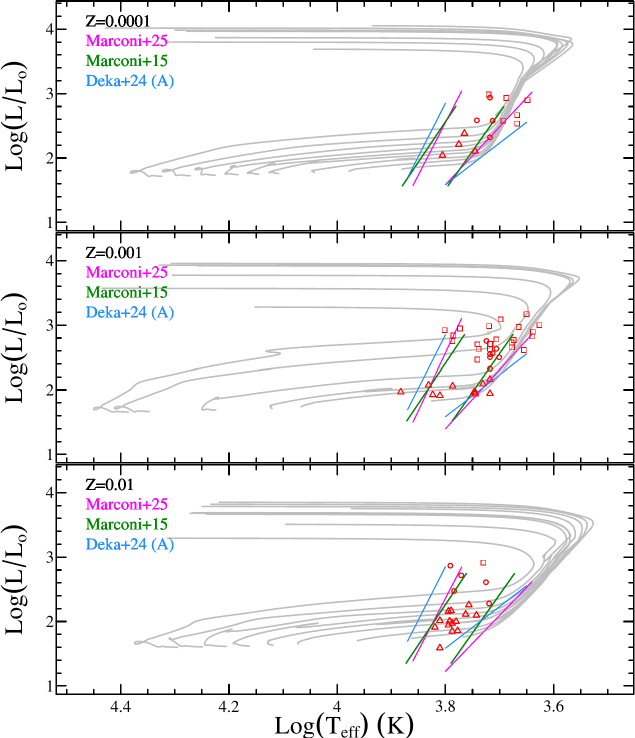}
    \caption{\label{is} Comparison between our observed sample of T2Cs and models. From top to bottom, the panels show models at metallicities Z = 0.0001, 0.001, and 0.01 and subsamples of our data for $\mathrm{[Fe/H]} < -1.30$ dex, $-1.30 < \mathrm{[Fe/H]} < -0.30$, and $\mathrm{[Fe/H]} > -0.30$ dex, respectively. Grey solid lines represent stars entering the instability strip at luminosities brighter than the Zero Age Horizontal Branch in old stellar populations \citep[][]{bono20}. The masses of the evolutionary tracks plotted are: for Z = 0.0001 M=[0.53, 0.54, 0.55, 0.56, 0.58, 0.60. 0.62, 0.65] $M_\odot$; for Z = 0.001 M=[0.505, 0.51, 0.53, 0.55, 0.56, 0.60] $M_\odot$; for Z = 0.01 M=[0.495, 0.50, 0.505, 0.51, 0.515, 0.52, 0.53, 0.54, 0.55] $M_\odot$. Green, blue, and magenta solid lines indicate the boundaries of the RR Lyrae instability strip \citep[extrapolated from][]{marconi2015}, the BLHer instability strip \citep{Deka2024}, and the T2C instability strip \citep{marconi2025}, calculated for pulsators up to P<10 d, respectively. Red triangles, circles and squares denote BLHer, WVir stars with P < 10 d, and WVir stars with P > 10 d, respectively. } 
        \end{figure}

The determination of the atmospheric parameters and iron abundances, combined with the \textit{Gaia} time-series photometry, allowed us to estimate the bolometric magnitudes and in turn the intrinsic luminosities of the stars at the phases corresponding to the spectroscopic observations. We apply the bolometric corrections by \citet[][]{chen2019} following the same procedure exploited in \citet[][]{marconi2021}. The resulting bolometric luminosities, together with the $T_{\rm eff}$, were used to populate the Hertzsprung--Russell diagram and to compare the observed distribution of T2Cs with theoretical instability-strip (IS) boundaries from \citet[e.g.][]{marconi2015,Deka2024}, and from \citet[][]{marconi2025}. This comparison is displayed in Fig.~\ref{is}. 
Since the predicted IS have been calculated for different metallicities, we decided to divide our sample in metallicity bins to ease the comparison. Moreover, in \citet[][]{marconi2025}, the model predictions stop at the pulsation period of 10 d. For this reason, in Fig.~\ref{is} we used triangles, circles, and squares to distinguish among BLHer, WVir stars with P < 10 d, and WVir stars with P > 10 d, respectively.

Figure~\ref{is} provides several interesting insights: 

\begin{itemize}
    \item For $\mathrm{[Fe/H]} < -1.30$ dex (top panel), the IS boundaries for T2Cs with $P < 10$ d predicted by \citet[][]{marconi2025} are able to reproduce the behaviour of BLHer and WVir stars with $4 < P < 20$ d. The BLHer IS boundaries \citep[][]{Deka2024} and the extrapolated RRL IS boundaries \citep[][]{marconi2015}, as predicted by \citet[][]{caputo2004bright,dicriscienzo07}, are consistent with the observed BLHer stars. 
    
    \item For $-1.30 < \mathrm{[Fe/H]} < -0.30$ dex (middle panel), all three IS predictions are consistent with the observations of BLHer and WVir stars with $P < 10$ d, whereas they show a wider spread in temperature than that displayed by the models. The comparison with the evolutionary tracks also clearly highlights the different evolutionary stages of BLHer and WVir stars.
    
    \item For $\mathrm{[Fe/H]} > -0.30$ dex (bottom panel), the IS predictions are consistent with the observed stars.
    It also appears that T2Cs with periods shorter than 10 days are more common at higher metallicities.
\end{itemize}
Even though the temperatures were derived at random pulsational phase for the observed sample, the overall agreement with theoretical predictions is impressive. More quantitative results could be obtained by comparing theoretical and empirical mean temperatures obtained from effective temperature curves.

\section{Metallicity dependence of Period-Wesenheit relations} \label{metdep}
Having derived homogeneous iron abundances for our targets, we are now in a position to investigate the metallicity dependence of the T2C PWZ relations. Our dataset approximately doubles the number of galactic T2Cs with spectroscopic abundance determinations available in the literature \citep[][]{maas2007, luck2011, Kovtyukh2018}, providing the largest homogeneous sample currently available for this purpose.
The PWZ relations were calibrated in the form:
\begin{equation}
    W= \alpha_p + \beta~( \log~P - \log~P_0) + \gamma ~ [Fe/H],
    \label{pwzeq}
\end{equation}
where $P_0$ is a pivotal period set to 12 days. However, for ease of use in subsequent analyses, in Tables~\ref{tab:pw_final} and \ref{tab:pw_final_2}  the coefficients are presented in the form 
   $ W= \alpha + \beta~ log~P + \gamma ~ [Fe/H]$,
where $\alpha = \alpha_P - \beta~log~P_0 $.

\begin{table}[!ht]
\centering
\renewcommand{\arraystretch}{1.3}
\caption{PWZ coefficients for the combined BLHer+WVir sample using different reddening laws.}
\label{tab:pw_final}
\begin{tabular}{lcccc}
\toprule
Relation & $\alpha$ & $\beta$ & $\gamma$ & $N$ \\
PW & {mag} & {mag} & {mag/dex} & \\
(1) & (2) & (3) & (4) & (5) \\
\midrule
\multicolumn{5}{c}{\textbf{BLHer + WVir}}\\
\midrule

Gaia G23 &
$-0.96^{+0.06}_{-0.05}$ &
$-2.36^{+0.08}_{-0.08}$ &
$-0.01^{+0.06}_{-0.07}$ &
74 \\

\rowcolor{lightgray}
Gaia R19 &
$-1.15^{+0.06}_{-0.05}$ &
$-2.46^{+0.08}_{-0.08}$ &
$-0.02^{+0.06}_{-0.06}$ &
74 \\

\rowcolor{lightgray}
VI C89 &
$-1.19^{+0.05}_{-0.04}$ &
$-2.49^{+0.07}_{-0.06}$ &
$+0.03^{+0.05}_{-0.05}$ &
72 \\

VI F99 &
$-1.11^{+0.05}_{-0.05}$ &
$-2.41^{+0.08}_{-0.08}$ &
$-0.01^{+0.06}_{-0.06}$ &
74 \\

VI G23 &
$-1.08^{+0.05}_{-0.05}$ &
$-2.39^{+0.08}_{-0.08}$ &
$-0.02^{+0.06}_{-0.06}$ &
74 \\

\rowcolor{lightgray}
VK C89 &
$-1.21^{+0.05}_{-0.05}$ &
$-2.44^{+0.09}_{-0.09}$ &
$+0.13^{+0.06}_{-0.07}$ &
74 \\

VK F99 &
$-1.20^{+0.05}_{-0.05}$ &
$-2.43^{+0.09}_{-0.09}$ &
$+0.13^{+0.06}_{-0.07}$ &
74 \\

VK G23 &
$-1.15^{+0.05}_{-0.05}$ &
$-2.39^{+0.09}_{-0.09}$ &
$+0.13^{+0.06}_{-0.06}$ &
74 \\

\rowcolor{lightgray}
JK C89 &
$-1.25^{+0.06}_{-0.05}$ &
$-2.42^{+0.09}_{-0.08}$ &
$+0.17^{+0.07}_{-0.07}$ &
74 \\

JK F99 &
$-1.27^{+0.06}_{-0.05}$ &
$-2.43^{+0.09}_{-0.09}$ &
$+0.17^{+0.07}_{-0.07}$ &
74 \\

JK G23 &
$-1.20^{+0.06}_{-0.05}$ &
$-2.40^{+0.08}_{-0.08}$ &
$+0.16^{+0.06}_{-0.07}$ &
74 \\

\bottomrule
\end{tabular}

\tablefoot{The PWZ relations have the form
$M_W = \alpha + \beta\log P + \gamma\,[\mathrm{Fe/H}]$ and have been
calibrated using a parallax zero-point counter-correction of
$-14\,\mu$as \citep[][]{Riess2021}.
The PWZ coefficients derived separately for the  BLHer and WVir samples are reported in
Table~\ref{tab:pw_final_2}.
The columns report: (1) the adopted PWZ relation, where R19, C89, F99,
and G23 refer, respectively, to
\citet[][]{Ripepi2019_reclassificaiton,cardelli1989relationship,
Fitzpatrick1999,gordon2023}; (2), (3), and (4) the $\alpha$, $\beta$,
and $\gamma$ coefficients and their uncertainties; and (5) the number
of stars used in the fit.}
\end{table}
Since $Gaia$ directly measures the parallaxes, the fit was performed in parallax space rather than by inverting the observed parallaxes into absolute magnitudes\footnote{Table \ref{data} lists the absolute Wesenheit magnitude of our sample calculated following the definitions of Table\ref{tab:coeffwes} and the corrected $Gaia$ parallaxes.}. For a given set of parameters, the predicted photometric parallax of each star is
\begin{equation}
    \varpi_{phot} = 10^{[-0.2 (w_{obs} - W_{model} - 10)]},
\end{equation}
where $ W_{model}$ comes from Eq.~\ref{pwzeq} and $ w_{obs} = m_{\lambda_1} - R_{\lambda_2,\lambda_1}~ (m_{\lambda_2} - m_{\lambda_1})$. According to the definition of Wesenheit \citep[][]{Madore1982}, $ R_{\lambda_2,\lambda_1} = A_{\lambda_1} / (A_{\lambda_2} - A_{\lambda_1})$. 
The adopted $ R_{\lambda_2,\lambda_1}$, based on 3 different reddening laws \citep[][]{cardelli1989relationship, Fitzpatrick1999,gordon2023}, are listed in Table \ref{tab:coeffwes}, according to \citet[][]{skowron2026}. We also report the empirical Wesenheit coefficient calculated by \citet[][]{Ripepi2019_reclassificaiton}.

\begin{table}[ht]
\centering
\footnotesize\setlength{\tabcolsep}{3pt} 
\caption{Definition of the Wesenheit magnitudes adopted in this work.}
  \label{tab:coeffwes}
\begin{tabular}{ll}
\toprule
Ref. & Wesenheit magnitude \\
\midrule
\citet[][]{cardelli1989relationship} 
& $W_{VI}=I-1.55(V-I)$ \\
& $W_{VK}=K-0.13(V-K)$ \\
& $W_{JK}=K-0.69(J-K)$ \\

\addlinespace

\citet[][]{Fitzpatrick1999} 
& $W_{VI}=I-1.39(V-I)$ \\
& $W_{VK}=Ks-0.12(V-K)$ \\
& $W_{JK}=K-0.74(J-K)$ \\

\addlinespace

\citet[][]{gordon2023}
& $W_{VI}=I-1.34(V-I)$ \\
& $W_G=G-1.67(G_{\rm BP}-G_{\rm RP})$ \\
& $W_{VK}=K-0.09(V-K)$ \\
& $W_{JK}=K-0.55(J-K)$ \\
\addlinespace

\citet[][]{Ripepi2019_reclassificaiton} 
& $W_G=G-1.90(G_{\rm BP}-G_{\rm RP})$ \\
\bottomrule
\end{tabular}
\end{table}

The fit was performed using Orthogonal Distance Regression (ODR), which accounts for uncertainties in both the response variable and the explanatory coordinates. In the present formulation, the observed coordinates are
$x_i = (w_i,\log P_i,{\rm [Fe/H]}_i)$,
and the fitted response variable is
$y_i = \varpi_{{\rm corr},i} + d $,
where $\varpi_{{\rm corr},i}$ are the $Gaia$ parallaxes already corrected for the parallax bias according to \citet{Lindegren2021} and \(d\) is the adopted parallax zero-point counter-correction \citep[see e.g.][and references therein]{ripepi2026}. In our analysis we choose the value $d=-14$ $\mu$as, found by \citet[][]{Riess2021} using a sample of Galactic CCs. 
Given the size of our sample and the strong degeneracy between the PW zero point and the parallax zero-point counter-correction, we do not attempt to fit both quantities simultaneously. Instead, we fix the counter-correction to $d=-14$ $\mu$as and fit only the zero point of the PW relation. This choice is also consistent with our previous analyses of Population II distance indicators \citep[S24,][]{sicignano2025}. We verified that varying the parallax zero-point counter-correction from 0 to $-$22 $\mu$as results in changes to the metallicity term that remain within 1 $\sigma$ (see Appendix~\ref{app:zp}).

ODR minimises the weighted distance between the observed data points and the model manifold. In simplified notation, the minimised quantity is

\begin{equation}
\chi^2_{\rm ODR} =
\sum_i \left[
\left(\frac{\Delta w_i}{\sigma_{w_{obs,i}   }}\right)^2
+
\left(\frac{\Delta \log P_i}{\sigma_{\log P,i}}\right)^2
+
\left(\frac{\Delta {\rm [Fe/H]}_i}{\sigma_{{\rm [Fe/H]},i}}\right)^2
+
\left(\frac{\Delta \varpi_i}{\sigma_{\varpi,i}}\right)^2
\right],
\end{equation}

subject to the model constraint
$\varpi_{{\rm obs},i}+\Delta \varpi_i =
10^{-0.2\left[(w_i+\Delta w_i)-W_i^{\ast}-10\right]}$,
where

\begin{equation}
W_i^{\ast}
=
\alpha_p
+
\beta\left[(\log P_i+\Delta \log P_i)-\log P_0\right]
+
\gamma\left({\rm [Fe/H]}_i+\Delta{\rm [Fe/H]}_i\right).
\end{equation}

The fitted physical parameters are therefore \(\alpha_p\), \(\beta\), and \(\gamma\). The quantities \(\Delta w_i\), \(\Delta \log P_i\), \(\Delta{\rm [Fe/H]}_i\), and \(\Delta\varpi_i\) are not additional physical parameters, but internal ODR coordinate corrections that minimise the total weighted distance.

The Gaia parallax variance was adopted as
$\sigma_{\varpi,i}^2 =
(1.1\,\sigma_{\varpi,{\rm Gaia},i})^2
+ (0.005\,{\rm mas})^2$ .

The uncertainty assigned to the Wesenheit coordinate in ODR was \[ \sigma_{w,{\rm ODR},i} = \sqrt{\sigma_{w,{\rm obs},i}^2+\sigma_{\rm int}^2}, \] where \(\sigma_{\rm int}\) is the intrinsic width of the instability strip. Since the fitted PWZ relation is a two-dimensional projection of the finite-width PLC relation \citep{sandage1958current}, an additional term accounting for the intrinsic scatter associated with the instability strip must be included in the uncertainty budget of the luminosity term \citep[e.g.][]{Riess2021, ripepi2026}.

To determine the most appropriate value of $\sigma_{\rm int}$ for our dataset, we explored a grid of intrinsic-width values and evaluated the quality of the resulting fits through a parallax-space $\chi^2$ analysis (the procedure and results are described in Appendix \ref{testchi}). This analysis led us to adopt $\sigma_{\rm int}$=0.15 mag for the Gaia-band PWZ relations, $\sigma_{\rm int}$ =0.12 mag for the VI and VK PWZ relations, and $\sigma_{\rm int}$=0.10 mag for the JK PWZ relation. 

Robust uncertainties on the fitted coefficients were estimated through bootstrap resampling. In each bootstrap realization, the stars were resampled with replacements, and \(w_{obs}\), \({\rm [Fe/H]}\), and parallax values were perturbed according to their measurement uncertainties. The relation was then refitted, and the quoted uncertainties were derived from the 16th and 84th percentiles of the bootstrap distributions.

\begin{figure}
\centering
    \includegraphics[width=10cm,clip,trim=0.5cm 0.5cm 0.5cm 0cm]{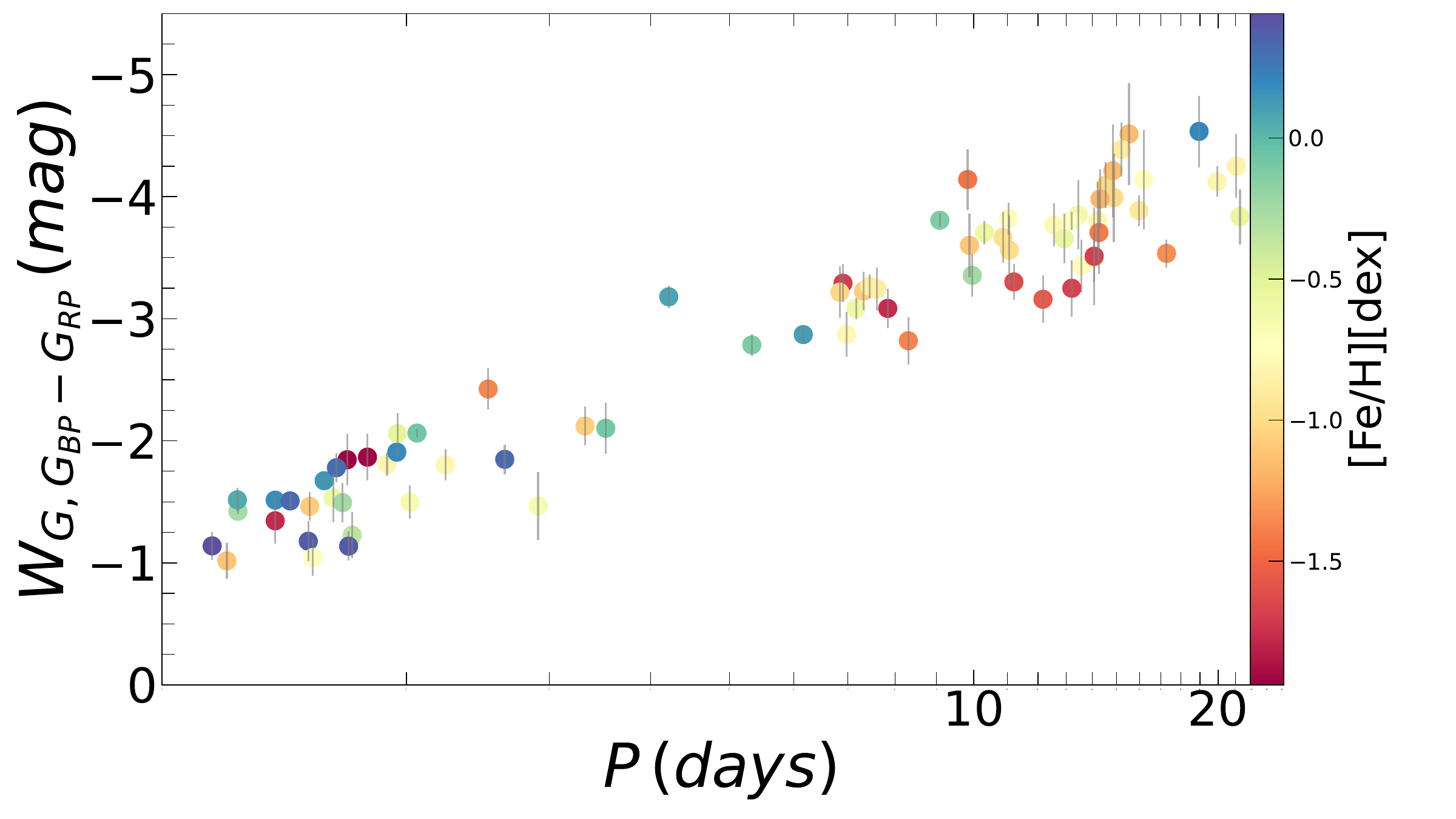}
    \caption{\label{pwgaia} PW relation in the the $Gaia$ bands in the magnitude space (the fit has been performed in parallax space). T2Cs are color coded according to their iron abundance. The data are listed in Table \ref{data}.} 
        \end{figure}

The results of the above outlined procedure are shown in Table~\ref{tab:pw_final} and \ref{tab:pw_final_2}, where we list the PWZ coefficients for the relations defined in Table \ref{tab:coeffwes}. The PWZ coefficients in  Table~\ref{tab:pw_final} have been calculated for BLHer and WVir separately and for both classes together. 
Fig.~\ref{pwgaia} shows visually the PW relation in the $Gaia$ bands, according to the \citet[][]{Ripepi2019_reclassificaiton} definition of Wesenheit, in the magnitude space.

Investigating Tables~\ref{tab:pw_final} and \ref{tab:pw_final_2} we note that i) the PWZ coefficients calculated separately for the BLHer and the WVir samples agree well within 1 $\sigma$, allowing us to use the complete sample of T2Cs for further analysis; ii) the PWZ coefficients for the same combination of filters but with Wesenheit magnitudes calculated using different reddening laws also agree within 1 $\sigma$.
On this basis, in the following, we adopt the results obtained with \citet{cardelli1989relationship} reddening law as baseline (shaded areas in Table~\ref{tab:pw_final}). 


Figures showing the covariance among the PWZ coefficients listed in Table \ref{tab:pw_final} are presented in Sect.~\ref{covariance} of the Appendix. In Table \ref{tab:pw_final}, we present the Galactic T2C PWZ relations based on the largest and most homogeneous sample currently available. The uncertainties and covariances of the PWZ coefficients are expected to decrease as samples spanning a wider metallicity range become available and as intensity-averaged mean magnitudes from optical (VI) and near-infrared time-series photometry are obtained.

\subsection{Comparison with theoretical and empirical literature PWZ relations} \label{sect:comparison}

The coefficients of the PWZ relations presented in Table~\ref{tab:pw_final} (adopting the results obtained with the \citet{cardelli1989relationship} reddening law) can be compared with various previous estimates based on pulsations models and observational studies of T2Cs as summarized in Table~\ref{tab:confronto} . 



\begin{itemize}
    \item In the $Gaia$ bands: $\alpha$ is negative by more than 2 $\sigma$ while $\beta$ is in perfect agreement with previous estimates from \citet[][]{marconi2025,lengen2025}, and S24; $\gamma$ is in agreement within 1-$\sigma$ with theoretical prediction from \citet[][]{marconi2025} and is smaller by 2-$\sigma$ compared to the one from \citet[][]{lengen2025} who assumed a mean metallicity value of [Fe/H]= -1.5 dex for the entire sample of T2Cs.
    \item In the VI bands: $\alpha$ agrees within 1-$\sigma$ when compared to \citet[][]{marconi2025}; S24 and is more negative by 2-$\sigma$ with \citet[][]{ngeow22}; $\beta$ and $\gamma$ agree within 1-$\sigma$ with values from \citet[][]{matsunaga2011,ngeow22,marconi2025} and S24.
    \item In the VKs bands: $\alpha$ is in agreement within 1-$\sigma$ with S24 but more negative by 2-$\sigma$ compared to \citet[][]{marconi2025}; $\beta$ is in agreement within 1-$\sigma$ with \citet[][]{ripepi2015, marconi2025}, and S24 but more negative by 2-$\sigma$ compared to \citet[][]{Bhardwaj2017_LMC}; $\gamma$ is in agreement within 1-$\sigma$ with \citet[][]{marconi2025}.
    \item In the JKs bands: $\alpha$ is in agreement within 1-$\sigma$ with \citet[][]{Wielgorski2022,ngeow22}, and S24 and within 2 $\sigma$ with \citet[][]{Wielgorski2022,marconi2025}; $\beta$ is in agreement within 1-$\sigma$ with all estimates in the previous literature; $\gamma$ is in agreement within 1-$\sigma$ with \citet[][]{ngeow22} and \citet[][]{marconi2025}and more positive by 2-$\sigma$ with \citet[][]{Wielgorski2022}.
\end{itemize}

\begin{figure*}
\centering
    \includegraphics[width=\textwidth]{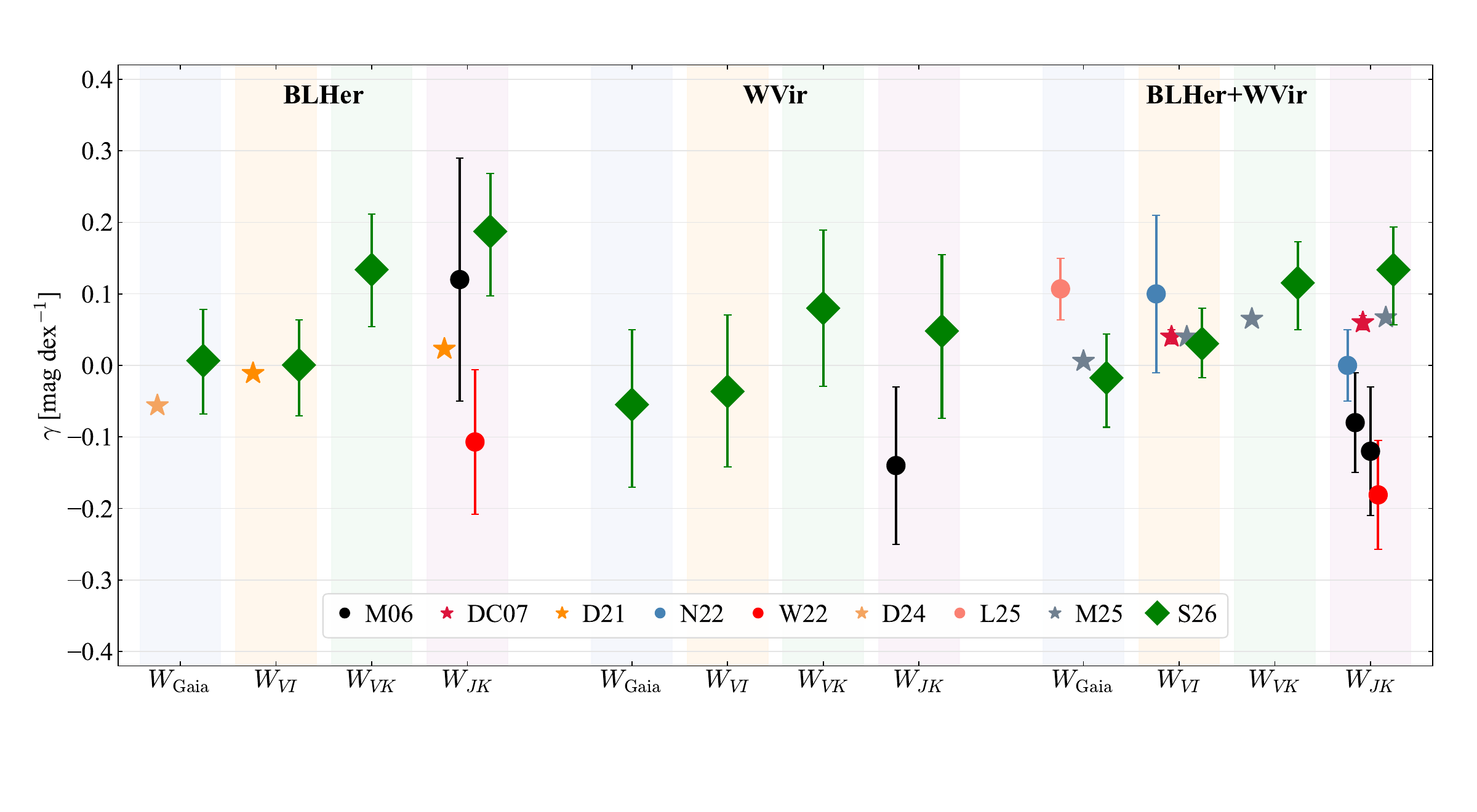}
    \caption{\label{gammacomp} Comparison of the $\gamma$ term and the relative uncertainty from the various PWZ obtained in this work (rombs) with theory (stars) and observations (circles): "M06" refers to \citet[][]{matsunaga2006}, "DC07" to \citet[][]{dicriscienzo07}, "D21" to \citet[][]{Das2021}, "N22" to \citet[][]{ngeow22}, "W22" to \citet[][]{Wielgorski2022}, "D24" to \citet[][]{das2024}, "L25" to \citet[][]{lengen2025}, "M25" to \citet[][]{marconi2025}, "S26" to this work. We note that \citet[][]{,matsunaga2006,ngeow22} $\gamma$ coefficients are taken from \citet[][]{das2026}.} 
        \end{figure*}

Since the main goal of this paper is the study of the metallicity dependence of T2Cs PWZ relations,  Fig. \ref{gammacomp} shows a comparison between our findings and those from the literature. Concentrating on the complete sample BLHer+WVir, the $\gamma$ coefficients are in excellent agreement with all theoretical predictions \citep[][]{dicriscienzo07,marconi2025} and the value by \citet{ngeow22} based on GGCs. On the contrary, a discrepancy larger than 1$\sigma$ is clearly visible compared with \citet{matsunaga2006} and especially \citet{Wielgorski2022}. These authors adopted a fitting PWZ method similar to the one presented in our paper, however their T2C sample is ten times smaller. 

\begin{table}
\footnotesize
\setlength{\tabcolsep}{4pt}
\caption{Comparison among published PW and PWZ relations for T2Cs.}
\label{tab:confronto}
\centering
\begin{adjustwidth}{-0.4cm}{-1cm}
\begin{tabular}{llllll}
\hline
PW &  & $\alpha$ & $\beta$ & $\gamma$ & Ref. \\
& & mag & mag & mag/dex & \\
(1) & (2) & {(3)} & (4) & (5) & (6) \\ 
\hline
Gaia & LMC & $-1.03 \pm 0.03$ & $-2.44 \pm 0.02$ &  & S24 \\
Gaia & Th. & $-0.992 \pm 0.003$ & $-2.466 \pm 0.006$ & $0.007 \pm 0.002$ & M25 \\
Gaia & MW & $-0.74 \pm 0.10$ & $-2.51 \pm 0.04$ & $0.11 \pm 0.04$ & L25 \\
\rowcolor{lightgray} Gaia & MW & $-1.15 \pm 0.05$ & $-2.46 \pm0.08$ & $-0.02\pm 0.07$ & S26 \\
\hline
VI & LMC &   & $-2.52 \pm 0.02$ &   & M11 \\
VI & GGCs &$ -0.99 \pm 0.08$ & $-2.43 \pm 0.07$ & $0.10 \pm 0.11$ & N22 \\
VI & LMC &$-1.14 \pm 0.03$ & $-2.49 \pm 0.02$ &   & S24 \\
VI & Th. & $-1.125 \pm 0.003$ & $-2.526 \pm 0.005$ & $0.040 \pm 0.002$ & M25 \\
\rowcolor{lightgray} VI & MW & $-1.19\pm0.05$ & $-2.49\pm0.07$ & $0.03 \pm 0.05$ & S26 \\
\hline
VK & LMC &   & $-2.49 \pm 0.03$ &   & R15 \\
VKs & LMC &   & $-2.28 \pm 0.04$ &   & B17 \\
VK & LMC & $-1.19 \pm 0.03$ & $-2.48 \pm 0.02$ &   &S24 \\ 
VK & Th. & $-1.024 \pm 0.003$ & $-2.567 \pm 0.006$ & $0.065 \pm 0.002$ & M25 \\
\rowcolor{lightgray} VK & MW & $-1.19\pm0.06$ & $-2.41 \pm 0.08$ & $0.12 \pm 0.07$ & S26 \\
\hline 
JK & LMC &   & $-2.52 \pm 0.03$ &   & R15 \\
JKs & LMC &   & $-2.35 \pm 0.06$ &   & B17 \\
JKs & GGCs &  $-1.28 \pm 0.03$ & $-2.46 \pm 0.03$ & $0.01 \pm 0.05$ & N22 \\
JK & LMC & $-1.17 \pm 0.01$ & $-2.54 \pm 0.03$  &   & W22 \\
JK & MW & $-1.34 \pm 0.02$ & $-2.40 \pm 0.10$ & $-0.18 \pm 0.08$ & W22 \\
JKs & LMC & $-1.23 \pm 0.03$ & $-2.50 \pm 0.02$ & & S24 \\
JK & Th. & $-1.084 \pm 0.003$ & $-2.627 \pm 0.006$ & $0.067 \pm 0.002$ & M25 \\
\rowcolor{lightgray} JK & MW & $-1.25\pm 0.06$ & $-2.42 \pm 0.09$ & $0.17\pm 0.08$ & S26 \\
\hline
\end{tabular}
\end{adjustwidth}
 \tablefoot{ Columns report: $(1)$ bands of the PW or PWZ relation; $(2)$ provenience of the sample;   $(3)$, $(4)$, $(5)$ the $\alpha$, the $\beta$, and, when present, the $\gamma$ coefficients with their uncertainties, respectively; $(6)$ reference:  "M11" refers to \cite{matsunaga2011}, "R15" refers to \cite{ripepi2015}, "B17" refers to \cite{Bhardwaj2017_LMC}, "W22" refers to \cite{Wielgorski2022}, "N22" refers to \cite{ngeow22}, "S24" refers to \cite{sicignano24}, "M25" refers to \cite{marconi2025}, "L25" refers to \cite{lengen2025}, "S26" (\textit{grey line}) refers to this work.}
\end{table}

\section{Applications: distances to GGCs}
\label{applications}

An obvious application of the PWZ relations derived in this work is the distance estimate of GGCs hosting T2Cs. Indeed, this class of well-studied objects allows us to both validate our relations against distances for the GGCs \citep[][]{Baumgardt2021} and to compare the T2Cs and RR Lyrae distance scales. 
 
To this aim, we collected a sample of 44 T2Cs belonging to 22 GGCs, taking from the literature only stars whose mean magnitudes were calculated as in our work \citep[i.e. as intensity-averaged mean magnitudes and periods, see][]{Bhardwaj2017_Bulge,bhar2021,braga2020separation}. We adopted distances (to compare with) from \citet{Baumgardt2021,bhardwaj2023,lengen2025},S24 and metallicity estimates from \citet[][]{carretta2009}. The $Gaia$ photometry was missing in the quoted publications so we added it from the $Gaia$ DR3 catalogue \citep[][]{Ripepi2023}. 

The distance moduli of the GGCs were determined through a linear least-squares formulation. For each T2C, the predicted absolute Wesenheit magnitude was computed from the adopted calibrated PWZ relation,
$W_i=\alpha_i+\beta_i\log P+\gamma_i{\rm [Fe/H]}_{\rm GC}$,
where $\alpha_i$, $\beta_i$, and $\gamma_i$ are the highlighted coefficients listed in Table~\ref{tab:pw_final} for our baseline results.
The observable quantity entering the fit is therefore:
$y_i=w_{{\rm obs},i}-W_i$,
which corresponds to an individual estimate of the distance modulus of the host GGC.
All measurements were assembled into the matrix equation :
$\mathbf{Y}=\mathbf{L}\mathbf{Q},$
where $\mathbf{Y}$ contains the values $y_i$, $\mathbf{Q}$ is the vector of unknown GGC dstance moduli, and $\mathbf{L}$ is a design matrix with one column per cluster. Each row of $\mathbf{L}$ contains a value of unity in the column corresponding to the cluster hosting the T2C and zeros elsewhere. This formulation allows all available PWZ relations to contribute simultaneously to the determination of a common distance modulus for each cluster, $\mu$.

For each measurement, the variance was computed as
\begin{equation}
\sigma_{y,i}^{2} = \sigma_{w,i}^{2} + \sigma_{\rm PW}^{2} + \left( \gamma_i,\sigma_{{\rm [Fe/H]},i} \right)^2,
\end{equation}
where $\sigma_{w,i}$ is the uncertainty of the observed Wesenheit magnitude, $\sigma_{\rm PWZ}$ is the intrinsic scatter of the adopted PWZ calibration, and $\sigma_{{\rm [Fe/H]},i}$ is the uncertainty on the cluster metallicity.

Because several PWZ relations were computed for the same star and therefore share common photometric measurements, the observational covariance matrix $\mathbf{C}$ was constructed including off-diagonal terms. For two PWZ measurements (a) and (b) derived from the same star,

\begin{equation}
{\rm Cov}(a,b) =\sum_{m} \frac{\partial w_a}{\partial m} \frac{\partial w_b}{\partial m} \sigma_m^2,
\end{equation}
where the sum runs over all photometric bands m entering the Wesenheit definitions and $\sigma_m$ is the corresponding photometric uncertainty. Consequently, measurements sharing one or more photometric bands are treated as correlated, while measurements from different stars are assumed to be independent.

The system was solved through generalised least-squares (GLS), yielding 
$\hat{\mathbf Q} = \left( \mathbf L^{T} \mathbf C^{-1} \mathbf L \right)^{-1} \mathbf L^{T} \mathbf C^{-1} \mathbf Y $,
with parameter covariance matrix
${\rm Cov}(\mathbf Q)= \left( \mathbf L^{T} \mathbf C^{-1} \mathbf L \right)^{-1}$.
The formal uncertainties of the GLS solution were obtained from the square root of the diagonal elements of {\rm Cov}$(\mathbf Q)$.
The adopted GGC distance modulus, $\mu$, was taken from the GLS solution, which corresponds to a weighted mean of the individual PWZ-based distance estimates, accounting for the full covariance matrix of the measurements. The adopted uncertainty was taken from the diagonal elements of the parameter covariance matrix.
As an additional diagnostic, we also computed the unweighted dispersion of the individual PWZ-based distance moduli for each cluster,
\begin{equation}
\sigma_{\rm PWZ} = \sqrt{ \frac{1}{N-1} \sum_{i=1}^{N} \left( \mu_i- \mu \right)^2},
\end{equation}
where \(N\) is the number of PWZ measurements available for the cluster.

\begin{figure}
\centering
    \includegraphics[width=8cm]{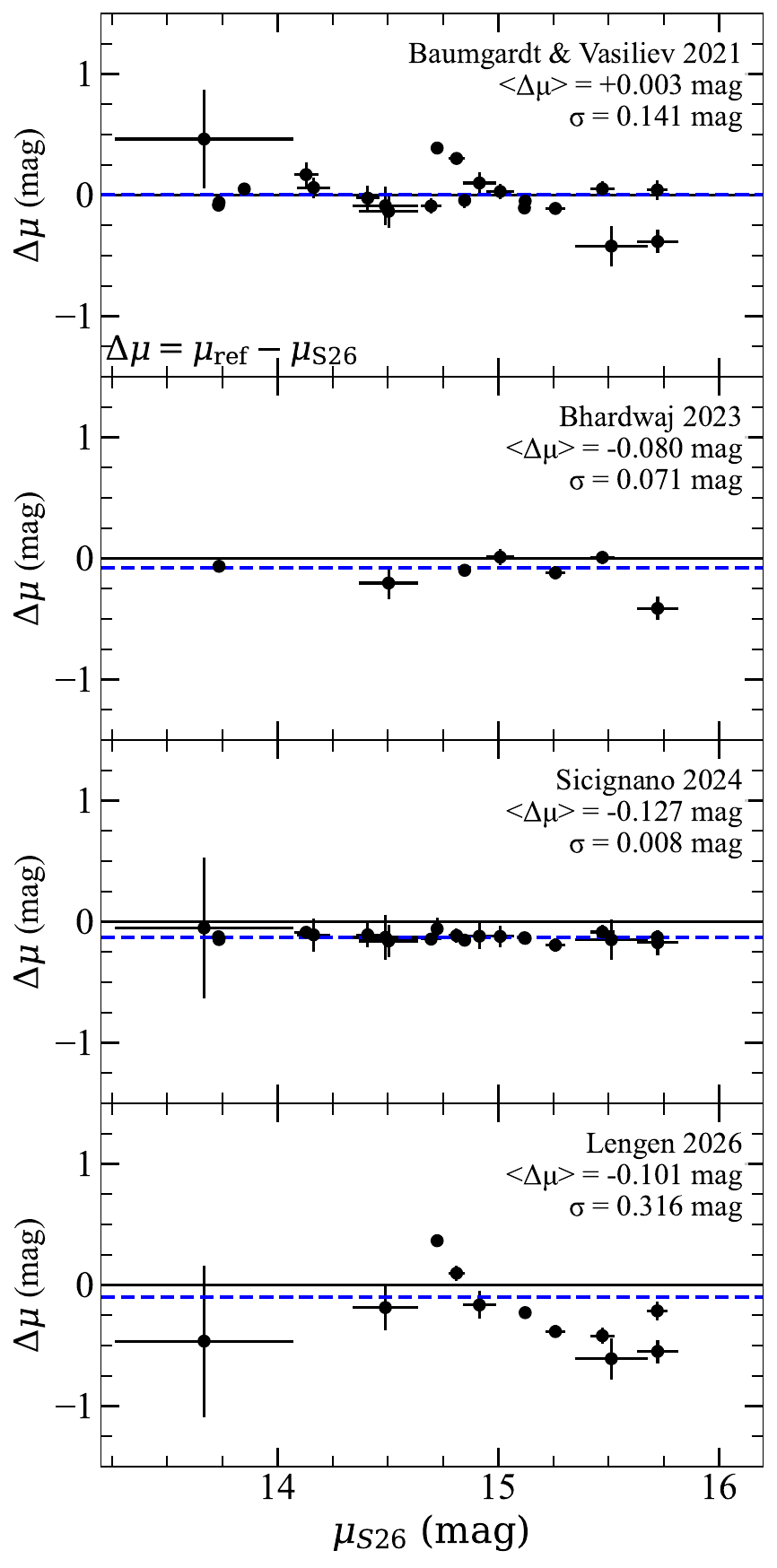}
    \caption{\label{ggcs} Comparison of the GGCs distances. $\langle\Delta\mu\rangle$ (dashed blue line) and $\sigma$ are the weighted mean and the dispersion of the various $\mu$. } 
        \end{figure}

The distances moduli of the GGCs based on T2Cs PWZ relations are listed in Table~\ref{tab:mu}. 
Figure~\ref{ggcs} shows the comparison between our GGCs distances and previous independent estimates by \citet{Baumgardt2021,bhardwaj2023,sicignano24,lengen2025}. Every panel of the figure reports also the relative weighted average of the distance moduli difference between the literature ("ref") and this work ("S26"): $\langle\Delta\mu\rangle= \mu_{ref}- \mu_{S26}$.
We note an excellent agreement between our results and previous GGC distance estimates by \citet[][]{Baumgardt2021}, based on an average of multiple methods, as well as with those by \citet[][]{bhardwaj2023}, which are based on PWZ relations for RRL variables. 
The current GGC distance moduli are approximately 0.1 mag larger than those reported in our previous work \citep[][]{sicignano24}, which relied on PW relations calibrated using the Large Magellanic Cloud (LMC) geometric distance \citet[][]{Pietrzynski_2019Natur.567..200P} and omitted the metallicity term. As noted in \citet[][]{sicignano24}, GGC distances derived from LMC-calibrated T2C PW relations were systematically smaller than the independent estimates provided by \citet[][]{Baumgardt2021}. By incorporating a metallicity term and calibrating the resulting PWZ relations with Gaia parallaxes, our newly derived distances now align well with these independent estimates. Furthermore, our GGC distances are larger than those recently derived by \citet[][]{lengen2025}. In their study, \citet[][]{lengen2025} observed that their distances were also smaller than those of \citet[][]{Baumgardt2021}, which they attributed to the averaging of heterogeneous and correlated calibration values. The discrepancy between our results and those of \citet[][]{lengen2025} primarily arises from the calibration approach: we anchored our PWZ relations using Gaia parallaxes alongside a global zero-point counter-correction of $-14$ $\mu$as, whereas their calibration relied on the Gaia parallaxes of the GGCs themselves. As noted already in Sect.\ref{sect:comparison}, this methodological divergence manifests directly as a difference in the derived $\alpha$ coefficient of the PWZ relations.

\begin{table}
\centering
\renewcommand{\arraystretch}{1.3}
\caption{Distances moduli to the GGCs based on T2Cs PWZ relations.}
\label{tab:mu}
\begin{tabular}{lcccc}
\toprule
GGC & $\mu$ & $\sigma_{\mu}$ & $\sigma_{PWZ}$ & $N$ \\
 & {mag} & {mag} & {mag} & \\ 
(1) & (2) & {(3)} & (4) & (5) \\ 
\midrule
HP1 & 14.16 & 0.07 & 0.15 & 4\\
  NGC2808 & 15.119 & 0.007 & 0.014& 4\\
  NGC5139 & 13.74 & 0.02 & 0.08& 28\\
  NGC5272 & 15.01 & 0.06 & 0.13& 4\\
  NGC5904 & 14.50$^+$ & 0.13 & 0.38& 8\\
  NGC5986 & 14.81$^+$ & 0.04 & 0.05& 2\\
  NGC6093 & 15.122 & 0.008 & 0.014& 3\\
  NGC6256 & 14.13 & 0.05 & 0.08& 2\\
  NGC6266 & 13.85 & 0.01 & 0.02& 4\\
  NGC6273 & 14.70 & 0.059 & 0.12& 6\\
  NGC6284 & 15.72 & 0.05 & 0.10& 4\\
  NGC6325 & 14.41 & 0.05 & 0.10& 4\\
  NGC6402 & 14.85 & 0.03 & 0.102& 14\\
  NGC6441 & 15.47 & 0.05 & 0.15& 8\\
  NGC6453 & 14.91$^+$ & 0.08 & 0.21& 8\\
  NGC6569 & 14.723$^+$ & 0.009 & 0.013& 2\\
  NGC6626 & 13.733 & 0.002 & 0.003& 2\\
  NGC6749 & 14.49 & 0.15 & 0.21& 2\\
  NGC6779 & 15.51$^+$ & 0.16 & 0.40& 6\\
  NGC7078 & 15.26 & 0.05 & 0.13& 8\\
  NGC7089 & 15.72$^+$ & 0.09 & 0.32& 12\\
  Terzan1 & 13.67 & 0.40 & 0.57& 2\\
\bottomrule
\end{tabular}
\tablefoot{ Columns meaning: $(1)$ GGC name; $(2)$ $\mu$ to the GGC accounting for the full covariance matrix of the measurements. The symbol $^+$ means that the periods of the T2Cs hosted by the specific GGCs is larger than our baseline periods; $(3)$ uncertainty on the $\mu$, derived as the square root of the diagonal elements of {\rm Cov}$(\mathbf Q)$; $(4)$ the unweighted dispersion of the individual PWZ-based distance moduli for each cluster; $(5)$ number of the individual PWZ-based distance moduli for each cluster.}
\end{table}

\section{Summary and Conclusions} \label{conclus}
In this tenth paper of the C-MetaLL series, we present a spectroscopic analysis of a sample of 74 MW T2Cs, providing the largest homogeneous spectroscopic sample currently available. The targets were selected among T2Cs with accurate and precise $Gaia$ parallaxes (relative errors < 10$\%$).

The available photometric data were partly gathered from the literature. In particular, $Gaia$ mean magnitudes in the $G$, $G_{BP}$, and $G_{RP}$ bands were available for the entire sample, whereas only a small fraction of the targets ($\sim 10\%$) had OGLE light-curve photometry in the $V$ and $I$ bands. In the near-infrared ($J$ and $K_s$) bands, only single-epoch 2MASS photometry was available. We therefore adopted the template technique \citep[][]{Ripepi2021, sicignano24} to estimate mean magnitudes. The lack of dedicated multi-epoch photometry remains one of the main limitations to the precision of PWZ coefficients determination, and, in turn, to their application to the extragalactic distance scale.

The stellar parameters and chemical abundances were derived through the EW method. The effective temperatures were derived by imposing excitation equilibium, while the surface gravities were estimated by imposing ionization equilibrium between Fe I and Fe II. The microturbulent velocity was derived by enforcing zero trend between iron abundance and line strength, as measured by the equivalent width. Final abundances were computed as the mean of the individual line abundances in the converged iteration. Our results are based on LTE plane parallel atmosphere models.

The comparison between the T2C abundances presented here and those from the literature was limited by the small overlap between the samples, as only ten targets have independent literature estimates. Overall, the metallicity distribution of our sample provides a good representation of the metallicity range expected for old stellar populations. 

The positions of our target stars within the instability strip agree well with theoretical predictions \citep[][]{marconi2025}.

The metallicity sample, together with the available photometric data, was used to determine the coefficients of the PWZ relations through an Orthogonal Distance Regression analysis. The Galactic PWZ relations were calibrated using Gaia parallaxes, applying the corrections by \citet[][]{Lindegren2021} and the counter zero-point offset of $-14~\mu$as as suggested by \citet[][]{Riess2021}. 
The comparison between our values of $\gamma$ and those from the literature should be treated with caution, since earlier studies relied either on small samples of field T2Cs or on metallicity measurements from GGCs. Noticeably, the $\gamma$ coefficients derived here agree with the theoretical predictions.

We applied our PWZ relations to GGCs hosting T2Cs, obtaining distance moduli which are in agreement with independent literature estimates. This promising result motivates us to further improve the precision and accuracy of T2Cs PWZ relations. To achieve this goal, we will expand the metallicity baseline by observing additional T2Cs, while exploiting homogeneous optical and near-infrared multi-band photometric datasets, some of which are already available through previous and ongoing observing campaigns at \href{http://www.rem.inaf.it/}{Rapid Eye Mount} at La Silla. In addition, the calibration of the zero points will be significantly improved thanks to the high-precision parallaxes expected from the forthcoming Gaia DR4 release.


Looking beyond the Milky Way, 4-metre Multi-Object Spectroscopic Telescope (\href{https://www.eso.org/sci/facilities/develop/instruments/4MOST.html}{4MOST}) will provide the opportunity to investigate T2C PWZ relations in the LMC, the ideal laboratory for measuring the width of the instability strip and further testing the metallicity effect on the coefficients of the PW relations in different stellar environments.

\begin{acknowledgements}
We thank the anonymous referee for their useful comments, which helped improve our manuscript. 
We used Generative Artificial Intelligence tools for coding assistance (e.g., creating figures). These tools are treated as consultants rather than content providers, and their suggestions are closely scrutinized and only accepted when deemed reliable by a human expert.
Based on observations collected at the European Southern Observatory under ESO programmes P115.28AS and P116.291H.
This research made use of TOPCAT \citep{Taylor2005}, NumPy \citep{Harris2020}, and Matplotlib \citep{Hunter2007}.
This research has used the SIMBAD database operated at CDS, Strasbourg, France and data from the European Space Agency (ESA) mission Gaia, processed by the Gaia Data Processing and Analysis Consortium. Funding for the DPAC has been provided by national institutions, in particular, the institutions participating in the Gaia Multilateral Agreement.  We acknowledge funding from: INAF GO-GTO grant 2023 “C-MetaLL - Cepheid metallicity in the Leavitt law” (P.I. V. Ripepi); PRIN MUR 2022 project (code 2022ARWP9C) 'Early Formation and Evolution of Bulge and Halo (EFEBHO),' PI: Marconi, M., funded by the European Union – Next Generation EU; Large Grant INAF 2023 MOVIE (P.I. M. Marconi). A.B. thanks the funding from the Anusandhan National Research Foundation (ANRF) under the Prime Minister Early Career Research Grant scheme (ANRF/ECRG/2024/000675/PMS). This research was supported by  the International Space Science Institute (ISSI) in Bern/Beijing through ISSI/ISSI-BJ International Team project ID \#24-603 - “EXPANDING Universe” (EXploiting Precision AstroNomical Distance INdicators in the Gaia Universe). G.D.S. acknowledges financial support from Gaia DPAC funds through INAF and the Agenzia Spaziale Italiana (ASI; contract 2025-10-HH.0, PI: M.G. Lattanzi).
T.S. and G.D.S. thank the Istituto Nazionale di Fisica Nucleare (INFN), Naples section, for support through specific initiatives QGSKY and Moonlight2. Finally, this paper is based on work supported by COST Action CA21136, "Addressing Observational Tensions in Cosmology with Systematics and Fundamental Physics (CosmoVerse)," funded by COST (European Cooperation in Science and Technology). 

\end{acknowledgements}

%
%

   \bibliographystyle{aa} 
   \bibliography{myBib} 

\newpage
\begin{appendix}
\section{Diagnostics on the abundance analysis} \label{diagnostics}
\begin{figure*}
\centering
    \includegraphics[width=17cm]{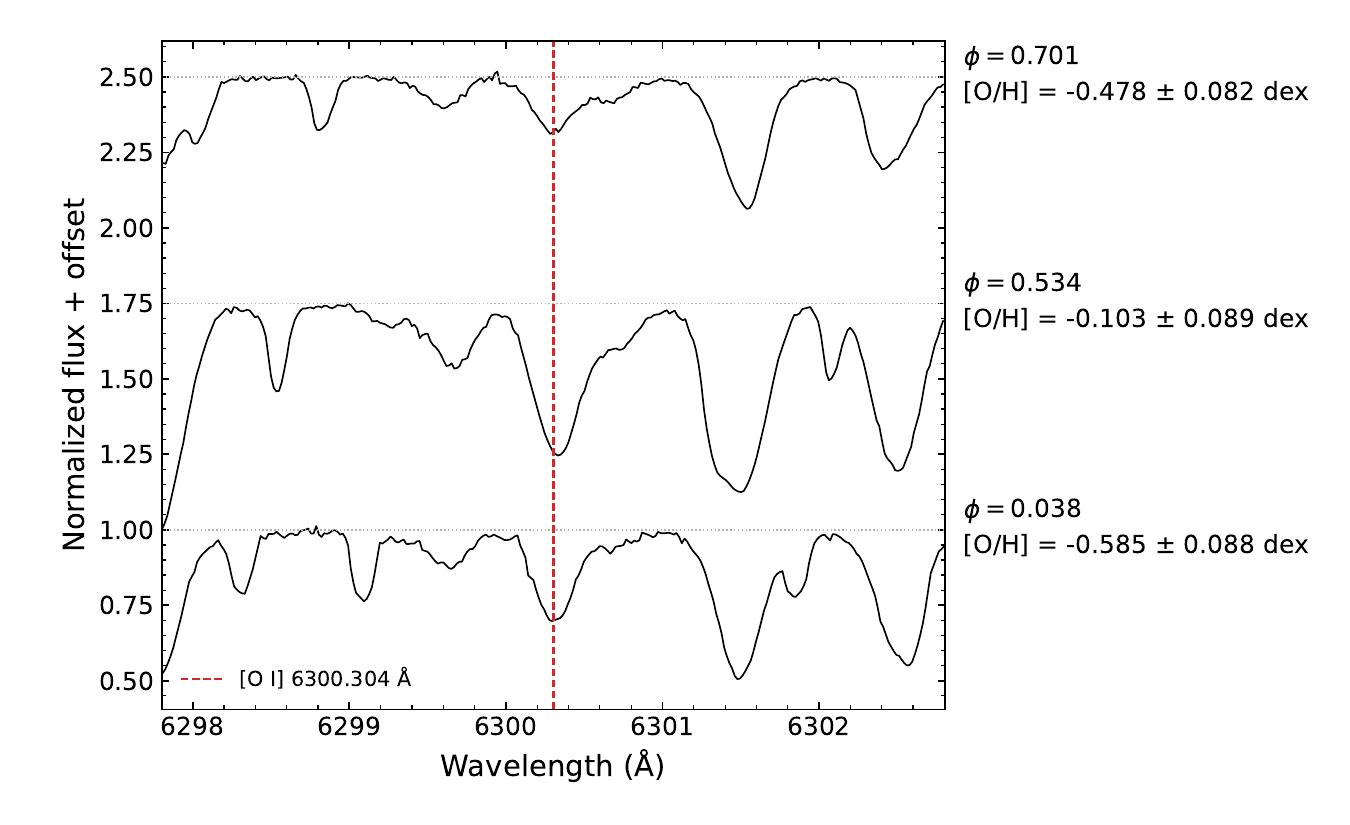}
    \caption{\label{oxygen} Oxygen line 6300.30\AA\ of different stars observed  different pulsational phases.}         \end{figure*}

Figure \ref{oxygen} shows the behavior of the [OI] line at 6300.30\AA\ for different pulsational phases.

\begin{figure*}
\centering
    \includegraphics[width=17cm]{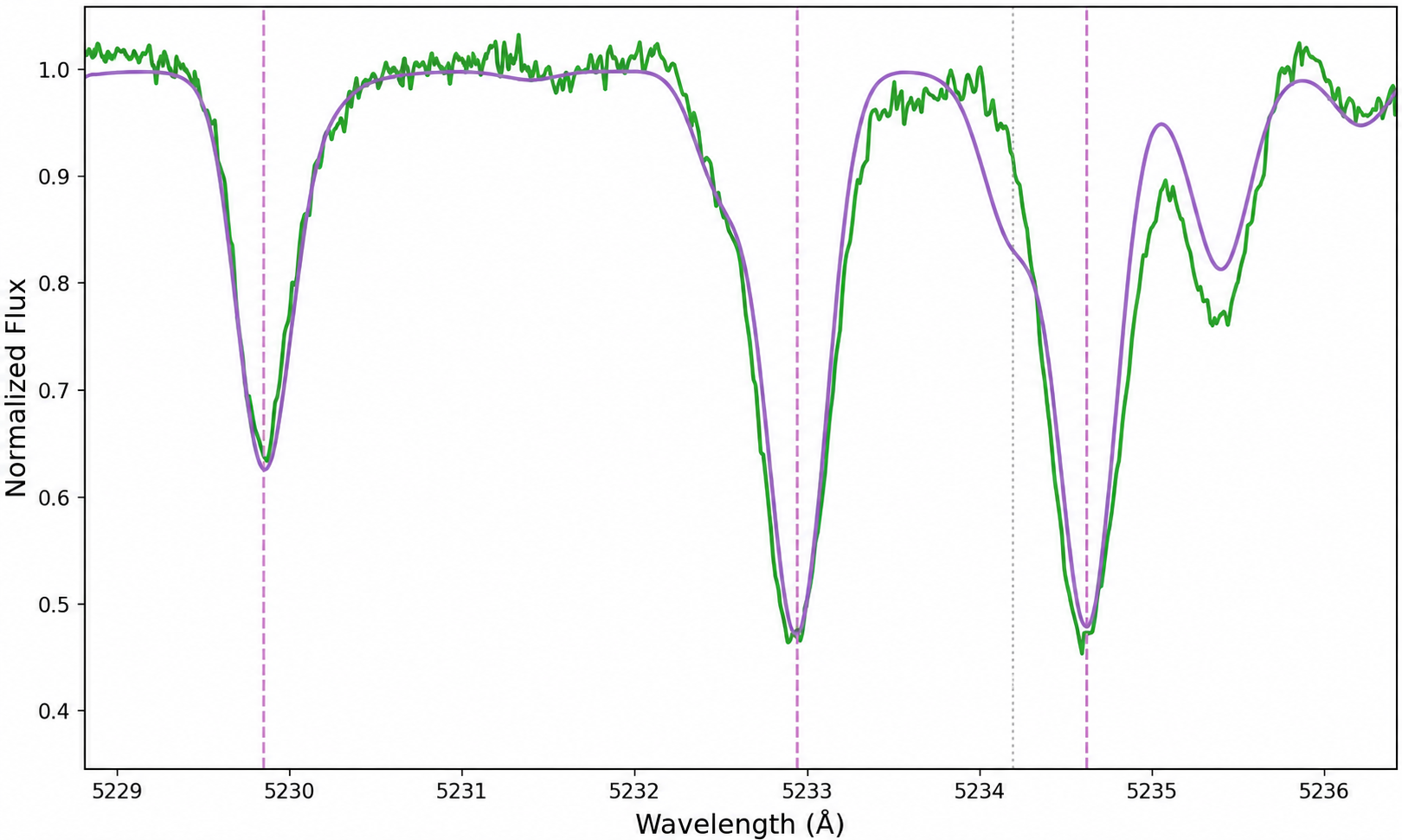}
    \caption{\label{spectra2} Example of an observed high-resolution spectrum (green) and a synthetic spectrum (purple) computed using the stellar parameters derived with GALA from EWs, over the wavelength range $\lambda \lambda$ 5229–5236\AA. Purple dashed and gray dotted vertical lines mark the positions of Fe I and Fe II lines, respectively.}        \end{figure*}

Figures \ref{spectra2} and \ref{spectra2} show examples of fitted lines in different regions of the spectral observations. In particular, ATLAS9 LTE plane parallel stellar atmosphere models are able to reproduce iron and oxygen lines.
\begin{figure*}
\centering
    \includegraphics[width=17cm]{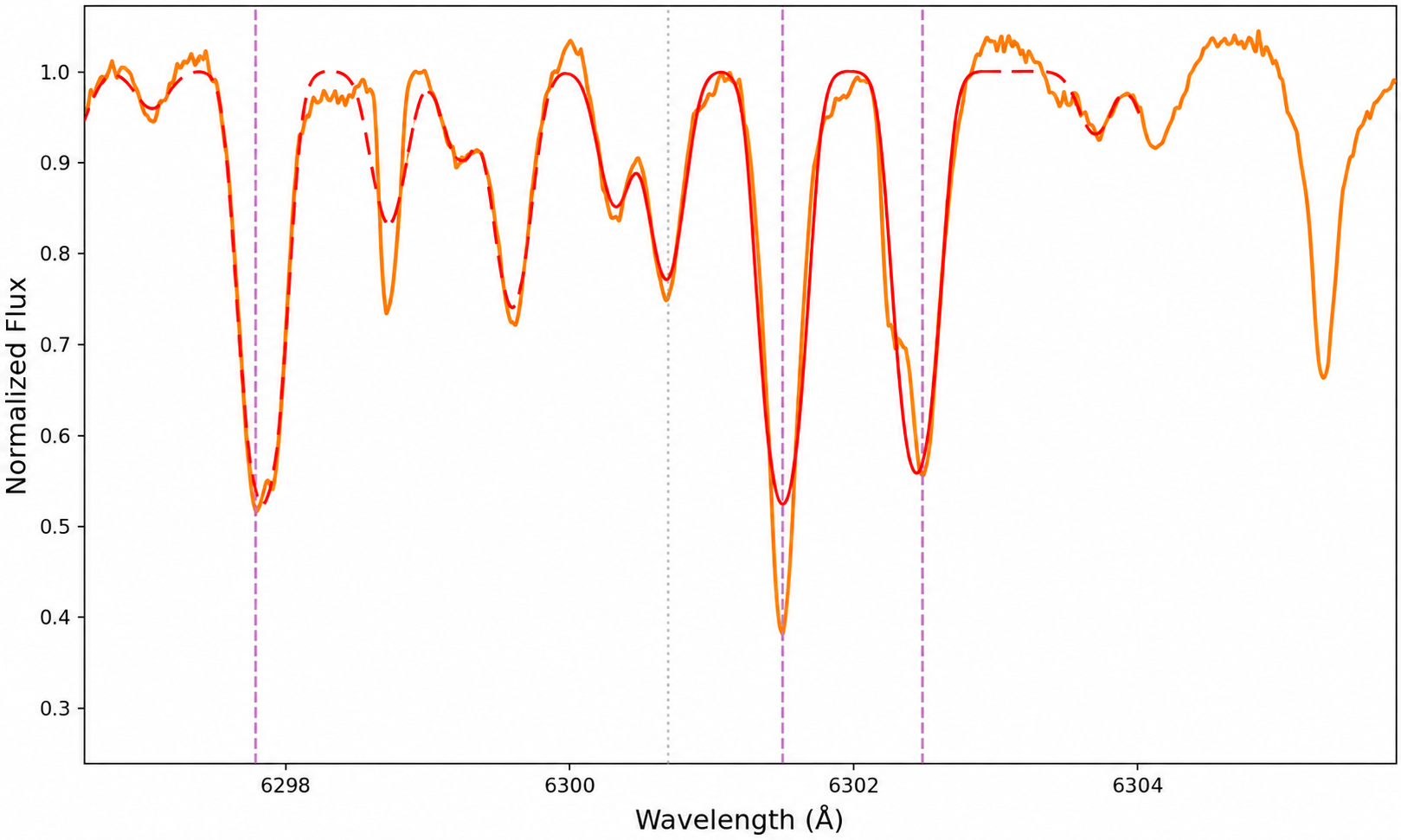}
    \caption{\label{spectra3} Example of an observed high-resolution spectrum (orange) and a synthetic spectrum (red) computed using the stellar parameters derived with GALA from EWs, over the wavelength range $\lambda \lambda$ 6297–6305\AA. Purple dashed and gray dotted vertical lines mark the positions of Fe I and Fe II lines, respectively.}         \end{figure*}

\begin{figure}
\centering
    \includegraphics[width=9cm]{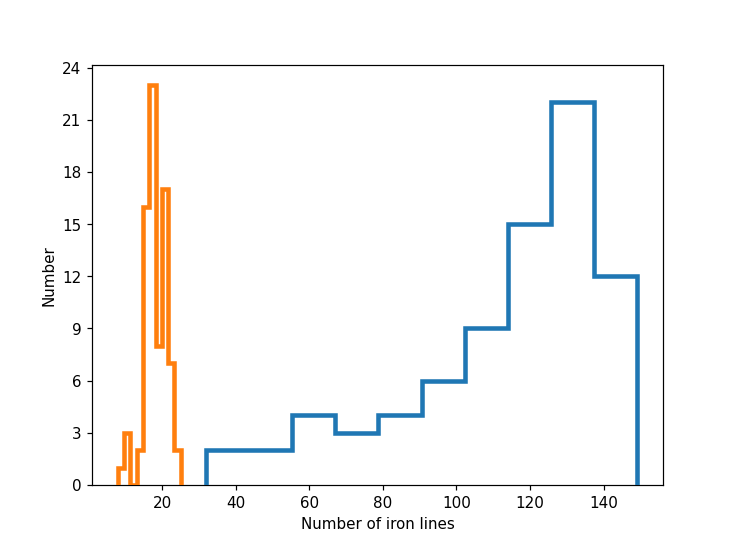}
    \caption{\label{nlines} Number of FeI (blue histogram) and FeII (orange histogram) spectral lines retained in the abundance analysis of each target after outlier rejection as part of the fitting procedure with the GALA code (see Section \ref{abund analysis}).} 
        \end{figure}

\begin{figure*}
\centering
    \includegraphics[width=19cm]{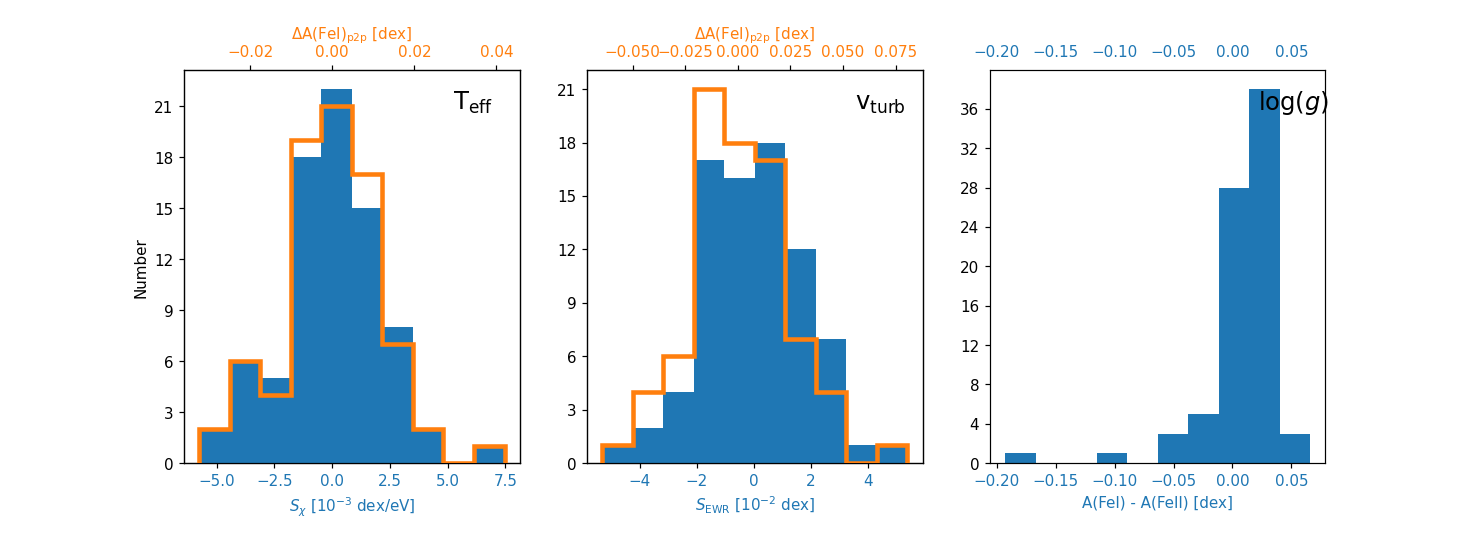}
    \caption{\label{check} Distributions of the final spectroscopic diagnostics obtained for the sample. Panel~(a): slope ($S_{\chi}$) of the Fe,\textsc{i} abundances as a function of the excitation potential ($\chi$), used to determine the effective temperature ($T_{\mathrm{eff}}$) (solid blue histogram; lower (x)-axis). The corresponding peak-to-peak abundance variations, ($\Delta A(\mathrm{Fe,\textsc{i}}){\mathrm{p2p}}$), calculated using Equation~\ref{deltaA}, are shown by the orange open histogram and refer to the upper (x)-axis. Panel~(b): slope ($S_{\mathrm{EWR}}$) of the Fe,\textsc{i} abundances as a function of the reduced equivalent width, ($\mathrm{EWR}=\log(EW/\lambda)$), used to constrain the microturbulent velocity ($v_{\mathrm{turb}}$) (solid blue histogram; lower (x)-axis). The associated values of ($\Delta A(\mathrm{Fe,\textsc{i}})_{\mathrm{p2p}}$) are represented by the orange open histogram and refer to the upper (x)-axis. Panel~(c): difference between the Fe,\textsc{i} and Fe,\textsc{ii} abundances, whose minimisation enforces ionisation constrains the surface gravity ($\log$ g).} 
        \end{figure*}

Table \ref{diffphas} reports the stellar parameters, iron and oxygen abundances, and pulsation phases at the time of observation. As expected, the stellar parameters vary throughout the pulsation cycle, whereas the derived chemical abundances remain consistent.
\begin{table*}
\begin{center}
\caption{Spectroscopic parameters and metal abundances of the three targets with two observations.\label{diffphas}}
\footnotesize\setlength{\tabcolsep}{3pt}
\begin{tabular}{rrrrrrrrr}
\hline
\noalign{\smallskip}
ID $Gaia$ & $\Phi_{\rm Obs}$ & $MJD_{\rm Obs}$ & SNR &
$T_{\rm eff}$ & $\xi$ & $\log\,g$ & [O/H] & [Fe/H] \\
& & d & @5870 \AA & K & km/s & dex & dex & dex \\
(1) & (2) & (3) & (4) & (5) & (6) & (7) & (8) & (9) \\
\noalign{\smallskip}
\hline
\noalign{\smallskip}
6045677196574103296 & 0.22 & 60845.17220 & 90 & 4750 $\pm$ 104 & 3.0 $\pm$ 0.11 & 0.3 $\pm$ 0.09 & $-$0.09 $\pm$ 0.13 & $-$0.55 $\pm$ 0.07 \\
6045677196574103296 & 0.15 & 60870.16993 & 51 & 4640 $\pm$ 103 & 3.1 $\pm$ 0.14 & 0.2 $\pm$ 0.07 & $-$0.11 $\pm$ 0.13 & $-$0.56 $\pm$ 0.08 \\
6638377559519795456 & 0.34 & 60870.32732 & 101 & 6390 $\pm$ 93  & 1.8 $\pm$ 0.21 & 1.4 $\pm$ 0.16 &  & $-$1.77 $\pm$ 0.06 \\
6638377559519795456 & 0.78 & 60872.31276 & 110 & 5660 $\pm$ 68  & 2.6 $\pm$ 0.12 & 0.9 $\pm$ 0.06 &  & $-$1.87 $\pm$ 0.05 \\
6706838857179992704 & 0.59 & 60845.31025 & 200 & 5220 $\pm$ 63  & 3.8 $\pm$ 0.14 & 0.4 $\pm$ 0.05 &  & $-$1.45 $\pm$ 0.04 \\
6706838857179992704 & 0.60 & 60855.15897 & 115 & 5250 $\pm$ 50  & 3.6 $\pm$ 0.13 & 0.5 $\pm$ 0.05 &  & $-$1.43 $\pm$ 0.03 \\
\noalign{\smallskip}
\hline
\noalign{\smallskip}
\end{tabular}

\tablefoot{Columns: $(1)$ Gaia source ID; $(2)$ phase of the observation;
$(3)$ Modified Julian Date of the observation; $(4)$ spectral
signal-to-noise ratio (SNR) at 5870 \AA; $(5)$ effective temperature;
$(6)$ microturbulent velocity; $(7)$ surface gravity; $(8)$ oxygen
abundance relative to the Sun, $[\mathrm{O/H}] \equiv
A(\mathrm{O})_{\rm star}-A(\mathrm{O})_{\odot}$, where
$A(\mathrm{O})=\log(N_{\mathrm O}/N_{\mathrm H})+12$ and
$A(\mathrm{O})_{\odot}=8.69$ \citep{2009asplund};
$(9)$ iron abundance relative to the Sun, $[\mathrm{Fe/H}] \equiv
A(\mathrm{Fe})_{\rm star}-A(\mathrm{Fe})_{\odot}$, where
$A(\mathrm{Fe})=\log(N_{\mathrm{Fe}}/N_{\mathrm H})+12$ and
$A(\mathrm{Fe})_{\odot}=7.50$ \citep{2009asplund}.}
\end{center}
\end{table*}
Figure \ref{nlines} shows the number of iron lines used in the analysis for each spectrum.
To test the reliability of the derived stellar parameters, Figure~\ref{check} shows the distributions of the diagnostics minimised during the fitting procedure. These include the slope \(S_{\chi}\) of the Fe\,\textsc{i} abundances as a function of the excitation potential \(\chi\), used to constrain \(T_{\mathrm{eff}}\); the slope \(S_{\mathrm{EWR}}\) of the Fe\,\textsc{i} abundances as a function of the reduced equivalent width, \(\mathrm{EWR}=\log(EW/\lambda)\), used to determine \(v_{\mathrm{turb}}\); and the difference between the Fe\,\textsc{i} and Fe\,\textsc{ii} abundances, adopted as the diagnostic for \(\log g\).

To quantify how the residual slopes affect the final iron abundances, panels~(a) and~(b) also show the corresponding peak-to-peak abundance variations. For each star, these are calculated as
\begin{equation} \label{deltaA}
    \Delta A(\mathrm{Fe\,\textsc{i}})_{\mathrm{p2p}}
    = S\,\ell ,
\end{equation}
where \(S\) denotes either \(S_{\chi}\) or \(S_{\mathrm{EWR}}\), while \(\ell\) is the range in \(\chi\) or EWR covered by the Fe\,\textsc{i} lines retained in the final fit. These quantities therefore measure the maximum abundance variation associated with the residual trends that remain after convergence. Their contribution is small, reaching at most only a few hundredths of a dex.


\section{PW data}
Table~\ref{data} presents the absolute Wesenheit magnitudes inferred from the corrected Gaia parallaxes. For each Wesenheit-index and reddening-law combination listed in Table~\ref{tab:coeffwes}, the observed Wesenheit magnitude was calculated as $ w_{obs} = m_{\lambda_1} - R_{\lambda_2,\lambda_1}~ (m_{\lambda_2} - m_{\lambda_1})$ 
The corresponding absolute Wesenheit magnitude was then obtained as
\begin{equation}
W_{\mathrm{abs}} = w_{\mathrm{obs}} - 5\log_{10}\left(\frac{1000} {\varpi_{\mathrm{corr}}+d}\right)+5,
\end{equation}
where the parallaxes are expressed in mas.

The ODR fit was performed in parallax space. The apparent Wesenheit magnitudes, $w_{\mathrm{obs}}$, were calculated
directly from the observed photometry. For a trial set of the PWZ
coefficients, the absolute Wesenheit magnitude of each star was
modelled as $M_{W,i}(\alpha_0,\beta,\gamma) =
\alpha_0+\beta(\log P_i-\log P_0)+\gamma[\mathrm{Fe/H}]_i$.

The corresponding photometric distance modulus was $\mu_{\mathrm{phot},i} = w_{\mathrm{obs},i}-M_{W,i}(\alpha_0,\beta,\gamma)$, from which the model photometric parallax was calculated as $\varpi_{\mathrm{phot},i}(\alpha_0,\beta,\gamma) = 10^{(10-\mu_{\mathrm{phot},i})/5}\ \mathrm{mas}$.

The PWZ coefficients $\alpha_0$, $\beta$, and $\gamma$ were determined through ODR by comparing the model photometric parallaxes with the corrected Gaia parallaxes, $\varpi_{\mathrm{obs},i} = \varpi_{\mathrm{Gaia},i}+d$.

\begin{table*}
\begin{center}

\caption{ $W_{\mathrm{abs}}$ of the target stars.}
\label{data}

\small
\setlength{\tabcolsep}{3pt}

\begin{tabular}{rrrrrrrrrrrr}
\hline
\noalign{\smallskip}
ID$_{\rm Gaia}$ & $P$ & $w_{G23}$ & $\sigma w_{G23}$ & $w_{G19}$ & $\sigma w_{G19}$ & $w_{JKC89}$ & $\sigma w_{JKC89}$ & $w_{JKF99}$ & $\sigma w_{JKF99}$ & $w_{JKG23}$ & $\sigma w_{JKG23}$ \\
 & days & mag & mag & mag & mag & mag & mag & mag & mag & mag & mag \\
(1) & (2) & (3) & (4) & (5) & (6) & (7) & (8) & (9) & (10) & (11) & (12) \\
\hline
\noalign{\smallskip}
892416 & 1.152 & $-1.05$ & 0.03 & $-1.22$ & 0.03 & $-1.10$ & 0.06 & $-1.12$ & 0.06 & $-1.06$ & 0.06 \\
141422 & 1.378 & $-1.30$ & 0.00 & $-1.53$ & 0.00 & $-1.61$ & 0.04 & $-1.63$ & 0.04 & $-1.55$ & 0.04 \\
803753 & 1.533 & $-0.88$ & 0.05 & $-1.13$ & 0.06 & $-1.22$ & 0.14 & $-1.24$ & 0.14 & $-1.15$ & 0.12 \\
239936 & 5.328 & $-2.59$ & 0.01 & $-2.86$ & 0.01 & $-3.07$ & 0.03 & $-3.10$ & 0.03 & $-2.99$ & 0.03 \\
922163 & 12.551 & $-3.56$ & 0.01 & $-3.95$ & 0.01 & $-4.02$ & 0.10 & $-4.06$ & 0.10 & $-3.91$ & 0.09 \\
\hline
\end{tabular}

\vspace{2mm}

\begin{tabular}{rrrrrrrrrrrr}
\hline
\noalign{\smallskip}
$w_{VIC89}$ & $\sigma w_{VIC89}$ & $w_{VIF99}$ & $\sigma w_{VIF99}$ & $w_{VIG23}$ & $\sigma w_{VIG23}$ & $w_{VKC89}$ & $\sigma w_{VKC89}$ & $w_{VKF99}$ & $\sigma w_{VKF99}$ & $w_{VKG23}$ & $\sigma w_{VKG23}$ \\
mag & mag & mag & mag & mag & mag & mag & mag & mag & mag & mag & mag \\
(13) & (14) & (15) & (16) & (17) & (18) & (19) & (20) & (21) & (22) & (23) & (24) \\
\hline
\noalign{\smallskip}
$-1.28$ & 0.09 & $-1.17$ & 0.09 & $-1.14$ & 0.08 & $-1.05$ & 0.04 & $-1.05$ & 0.04 & $-1.00$ & 0.04 \\
$-1.61$ & 0.09 & $-1.47$ & 0.08 & $-1.42$ & 0.08 & $-1.57$ & 0.03 & $-1.56$ & 0.03 & $-1.50$ & 0.02 \\
$-1.35$ & 0.06 & $-1.19$ & 0.06 & $-1.14$ & 0.06 & $-1.14$ & 0.09 & $-1.12$ & 0.09 & $-1.06$ & 0.09 \\
$-2.95$ & 0.09 & $-2.78$ & 0.08 & $-2.73$ & 0.08 & $-2.97$ & 0.02 & $-2.95$ & 0.02 & $-2.88$ & 0.02 \\
$-4.16$ & 0.06 & $-3.89$ & 0.06 & $-3.81$ & 0.05 & $-3.91$ & 0.06 & $-3.89$ & 0.06 & $-3.77$ & 0.06 \\
\hline
\end{tabular}

\tablefoot{ The table is split into two blocks for readability. The complete table, including the full Gaia identifiers, will be available in machine-readable format at the (\href{https://cds.unistra.fr/}{CDS}). Columns: (1) Gaia identifier. (2) Pulsation period, in days. (3)--(24) Absolute observed Wesenheit magnitudes and associated uncertainties for the corresponding photometric combinations listed in Table \ref{tab:coeffwes}.}

\end{center}
\end{table*}

\section{Validation of the Parallax Zero-Point Counter-Correction} \label{app:zp}
To assess the impact of the adopted parallax zero-point counter-correction on the calibration of the PWZ relations, we repeated the ODR fit by varying the counter-correction over the range from 0 to $-$ 22 $\mu${as} \citep[][]{Molinaro2023}. 
For each assumed counter-correction, we determined the coefficients of the PWZ relations while keeping the fitting procedure unchanged. We find that the metallicity coefficient increases with increasing absolute counter-correction. However, the corresponding variations remain within 1 $\sigma$, demonstrating that our results are robust against the adopted value of the parallax zero-point counter-correction.

\begin{figure}[H]
\centering
    \includegraphics[width=9cm,clip]{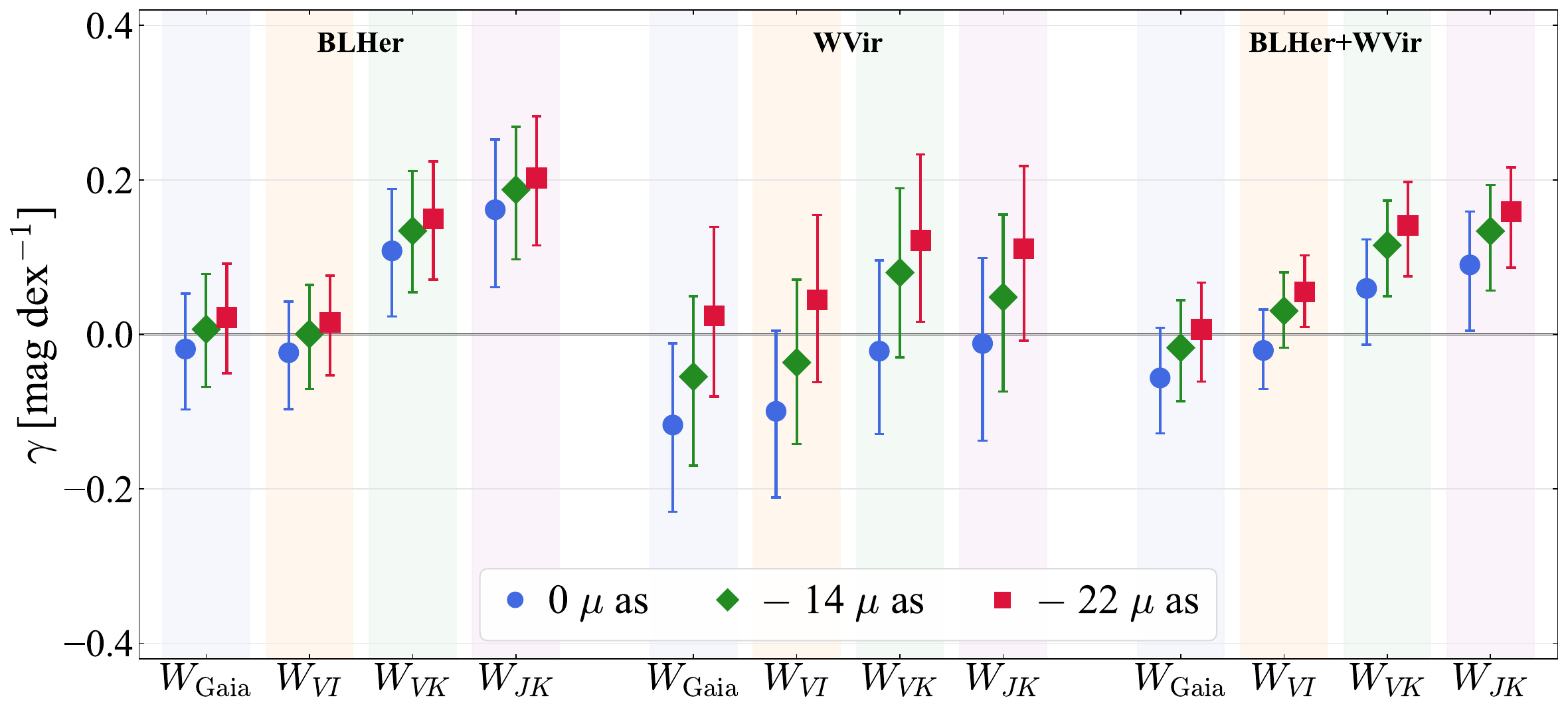}
    \caption{\label{3gamma} Comparison of the $\gamma$ term and the relative uncertainty adopting different parallax zero-point counter-corrections: blue circles, green diamonds, and red squares refer, respectively, to d = 0;  $-$ 14 $\mu$as \citep[][]{Riess2021}; $-$ 22 $\mu$as \citep[][]{Molinaro2023}.} 
        \end{figure}

These tests support the choice of adopting a fixed counter-correction of d = $-$ 14 $\mu$as throughout this work.
Fig. \ref{3gamma} shows the $\gamma$ coefficients and the relative uncertainties given a different parallax zero-point counter-correction.

\section{Test $\chi^2$} \label{testchi}
To determine the most appropriate value of the instability-strip width, we fitted the T2C PWZ relations in the Gaia, (JK), (VK), and (VI) bands (the latter two adopting the Cardelli reddening law). The fitting procedure follows that described by \citet{ripepi2026}.

A grid of values for the intrinsic width of the instability strip was explored and the quality of the resulting fits was evaluated through the reduced $\chi^2$ statistic. The adopted values of $\sigma_{\rm int}$ were selected as a compromise between obtaining an acceptable description of the observed scatter and remaining consistent with independent empirical estimates of the intrinsic dispersion of T2C PW relations \citet[][]{sicignano24}.

Even with the adopted values of $\sigma_{\rm int}$, the resulting reduced $\chi^2$ values remain typically around $\chi^2_{\rm red}\sim$ 1.4, indicating the presence of residual variance beyond that accounted for by the quoted observational uncertainties and the adopted intrinsic-width prescription. 
The $\chi^2$ statistic minimized in this analysis was computed in parallax space as
\begin{equation}
\chi^2 =
\sum_i
\frac{
\left(\varpi_{{\rm obs},i}-\varpi_{{\rm phot},i}\right)^2
}
{\sigma_{{\rm tot},i}^2},    
\end{equation}
where $\varpi_{\rm obs}$ is the Gaia parallax, $\varpi_{\rm phot}$ is the photometric parallax predicted by the PWZ relation, and $\sigma_{\rm tot}$ includes the contributions from the parallax uncertainty, the propagated photometric uncertainty, the metallicity uncertainty, and the adopted intrinsic width of the instability strip.
This is the same objective function underlying the ODR analysis in Sect. \ref{metdep}.
The coefficients obtained from the two methods are fully consistent within their uncertainties, indicating that the inferred PWZ relations are robust against the adopted fitting methodology.

The results of this analysis are summarized in Table~\ref{chisquare}.

\begin{table}[!ht]
\centering
\footnotesize
\setlength{\tabcolsep}{4pt} 
\caption{\label{chisquare}
PWZ coefficients obtained from the MCMC analysis for the combined
BLHer + WVir sample. Uncertainties correspond to the 16th and
84th percentiles of the posterior distributions.}
\begin{tabular}{lcccccc}
\hline
PW & $\sigma_{\rm int}$ &
$\alpha$ &
$\beta$ &
$\gamma$ &
$\chi^2_{\rm red}$ \\
& mag & mag & mag & mag/dex & \\
\hline

Gaia &
0.15 &
$-1.13\pm0.05$ &
$-2.45\pm0.07$ &
$-0.01\pm0.05$ &
1.47 \\

VI &
0.12 &
$-1.22\pm0.05$ &
$-2.48\pm0.07$ &
$-0.01\pm0.05$ &
1.48 \\
VK &
0.14 &
$-1.20\pm0.05$ &
$-2.42\pm0.07$ &
$+0.13\pm0.05$ &
1.55 \\
JK &
0.10 &
$-1.23\pm0.05$ &
$-2.41\pm0.08$ &
$+0.17\pm0.05$ &
1.50 \\

\hline
\end{tabular}
\end{table}
\section{PWZ coefficients and their covariance}\label{covariance}
The following plots show the covariance among PWZ cefficients of the higleted relations presented in Table \ref{tab:pw_final}.

\begin{table}[!ht]
\centering
\renewcommand{\arraystretch}{1.3}
\caption{PWZ coefficients for the BLHer and WVir samples using different reddening laws.}
\label{tab:pw_final_2}

\begin{tabular}{lcccc}
\toprule
Relation & $\alpha$ & $\beta$ & $\gamma$ & $N$ \\
PW & {mag} & {mag} & {mag/dex} & \\
(1) & (2) & (3) & (4) & (5) \\
\midrule

\multicolumn{5}{c}{\textbf{BLHer}}\\
\midrule

Gaia G23 &
$-0.92^{+0.10}_{-0.10}$ &
$-2.47^{+0.57}_{-0.40}$ &
$0.01^{+0.07}_{-0.07}$ &
30 \\

Gaia R19 &
$-1.07^{+0.11}_{-0.12}$ &
$-2.76^{+0.67}_{-0.46}$ &
$0.01^{+0.07}_{-0.07}$ &
30 \\

VI C89 &
$-1.09^{+0.10}_{-0.12}$ &
$-3.02^{+0.64}_{-0.42}$ &
$0.00^{+0.06}_{-0.06}$ &
30 \\

VI F99 &
$-1.01^{+0.10}_{-0.10}$ &
$-2.80^{+0.57}_{-0.37}$ &
$0.00^{+0.06}_{-0.06}$ &
30 \\

VI G23 &
$-0.99^{+0.09}_{-0.10}$ &
$-2.76^{+0.57}_{-0.35}$ &
$0.00^{+0.06}_{-0.06}$ &
30 \\

VK C89 &
$-1.15^{+0.09}_{-0.10}$ &
$-2.62^{+0.46}_{-0.38}$ &
$0.18^{+0.07}_{-0.07}$ &
30 \\

VK F99 &
$-1.15^{+0.09}_{-0.10}$ &
$-2.60^{+0.46}_{-0.38}$ &
$0.18^{+0.07}_{-0.07}$ &
30 \\

VK G23 &
$-1.12^{+0.09}_{-0.09}$ &
$-2.50^{+0.42}_{-0.37}$ &
$0.18^{+0.06}_{-0.06}$ &
30 \\

JK C89 &
$-1.24^{+0.09}_{-0.09}$ &
$-2.40^{+0.45}_{-0.41}$ &
$0.22^{+0.09}_{-0.08}$ &
30 \\

JK F99 &
$-1.25^{+0.09}_{-0.10}$ &
$-2.41^{+0.45}_{-0.42}$ &
$0.22^{+0.09}_{-0.09}$ &
30 \\

JK G23 &
$-1.18^{+0.11}_{-0.09}$ &
$-2.43^{+0.42}_{-0.43}$ &
$0.16^{+0.07}_{-0.07}$ &
29 \\

\midrule
\multicolumn{5}{c}{\textbf{WVir}}\\
\midrule

Gaia G23 &
$-1.19^{+0.25}_{-0.32}$ &
$-2.17^{+0.25}_{-0.27}$ &
$-0.06^{+0.11}_{-0.12}$ &
44 \\

Gaia R19 &
$-1.34^{+0.29}_{-0.30}$ &
$-2.30^{+0.26}_{-0.28}$ &
$-0.05^{+0.11}_{-0.11}$ &
44 \\

VI C89 &
$-1.38^{+0.30}_{-0.31}$ &
$-2.34^{+0.27}_{-0.31}$ &
$-0.04^{+0.10}_{-0.11}$ &
44 \\

VI F99 &
$-1.31^{+0.28}_{-0.31}$ &
$-2.24^{+0.25}_{-0.29}$ &
$-0.04^{+0.11}_{-0.10}$ &
44 \\

VI G23 &
$-1.28^{+0.27}_{-0.32}$ &
$-2.21^{+0.25}_{-0.28}$ &
$-0.04^{+0.11}_{-0.11}$ &
44 \\

VK C89 &
$-1.12^{+0.27}_{-0.35}$ &
$-2.58^{+0.32}_{-0.28}$ &
$0.05^{+0.12}_{-0.13}$ &
44 \\

VK F99 &
$-1.12^{+0.27}_{-0.35}$ &
$-2.57^{+0.32}_{-0.28}$ &
$0.04^{+0.12}_{-0.13}$ &
44 \\

VK G23 &
$-1.05^{+0.27}_{-0.33}$ &
$-2.54^{+0.31}_{-0.26}$ &
$0.06^{+0.12}_{-0.12}$ &
44 \\

JK C89 &
$-1.32^{+0.24}_{-0.30}$ &
$-2.42^{+0.26}_{-0.25}$ &
$0.08^{+0.10}_{-0.11}$ &
44 \\

JK F99 &
$-1.35^{+0.24}_{-0.29}$ &
$-2.42^{+0.26}_{-0.25}$ &
$0.08^{+0.10}_{-0.11}$ &
44 \\

JK G23 &
$-1.25^{+0.23}_{-0.31}$ &
$-2.42^{+0.27}_{-0.26}$ &
$0.08^{+0.10}_{-0.11}$ &
44 \\

\bottomrule
\end{tabular}

\tablefoot{The PWZ relations have the form
$M_W=\alpha+\beta\log P+\gamma\,[\mathrm{Fe/H}]$ and have been
calibrated using a parallax zero-point counter-correction of
$-14\,\mu$as \citep[][]{Riess2021}.
The columns report: (1) the adopted PWZ relation, where R19, C89, F99,
and G23 refer, respectively, to
\citet[][]{Ripepi2019_reclassificaiton,cardelli1989relationship,
Fitzpatrick1999,gordon2023}; (2), (3), and (4) the $\alpha$, $\beta$,
and $\gamma$ coefficients and their uncertainties; and (5) the number
of stars used in the fit.}
\end{table}
\begin{figure}
\centering
    \includegraphics[
    width=0.67\textwidth,
    trim=5.5cm 0.08cm 0cm 2cm,
    clip]{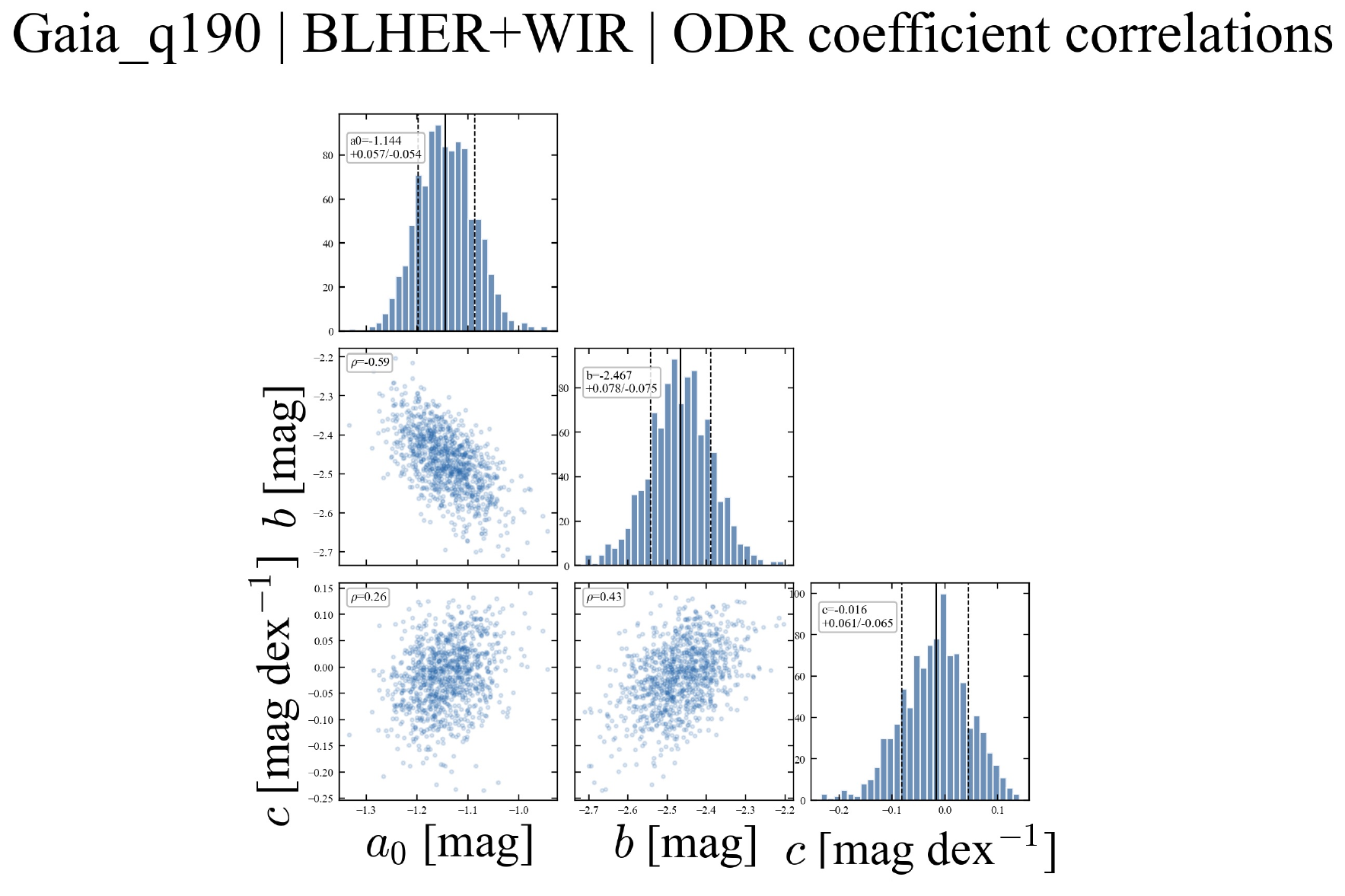}
    \caption{\label{cov 3} Covariance among coefficients of the PWZ relation for the BLHer and the WVir in Gaia bands, according to the Wesenheit definition of \citet[][]{Ripepi2019_reclassificaiton}. Points are ODR refits from bootstrap resampling; p is the Pearson coefficient. In each corner plot, the solid line marks the coefficient value, while the dashed lines indicate its uncertainty.} 
\end{figure}
\begin{figure}
\centering
    \includegraphics[
    width=0.67\textwidth,
    trim=5.5cm 0.08cm 0cm 2cm,
    clip]{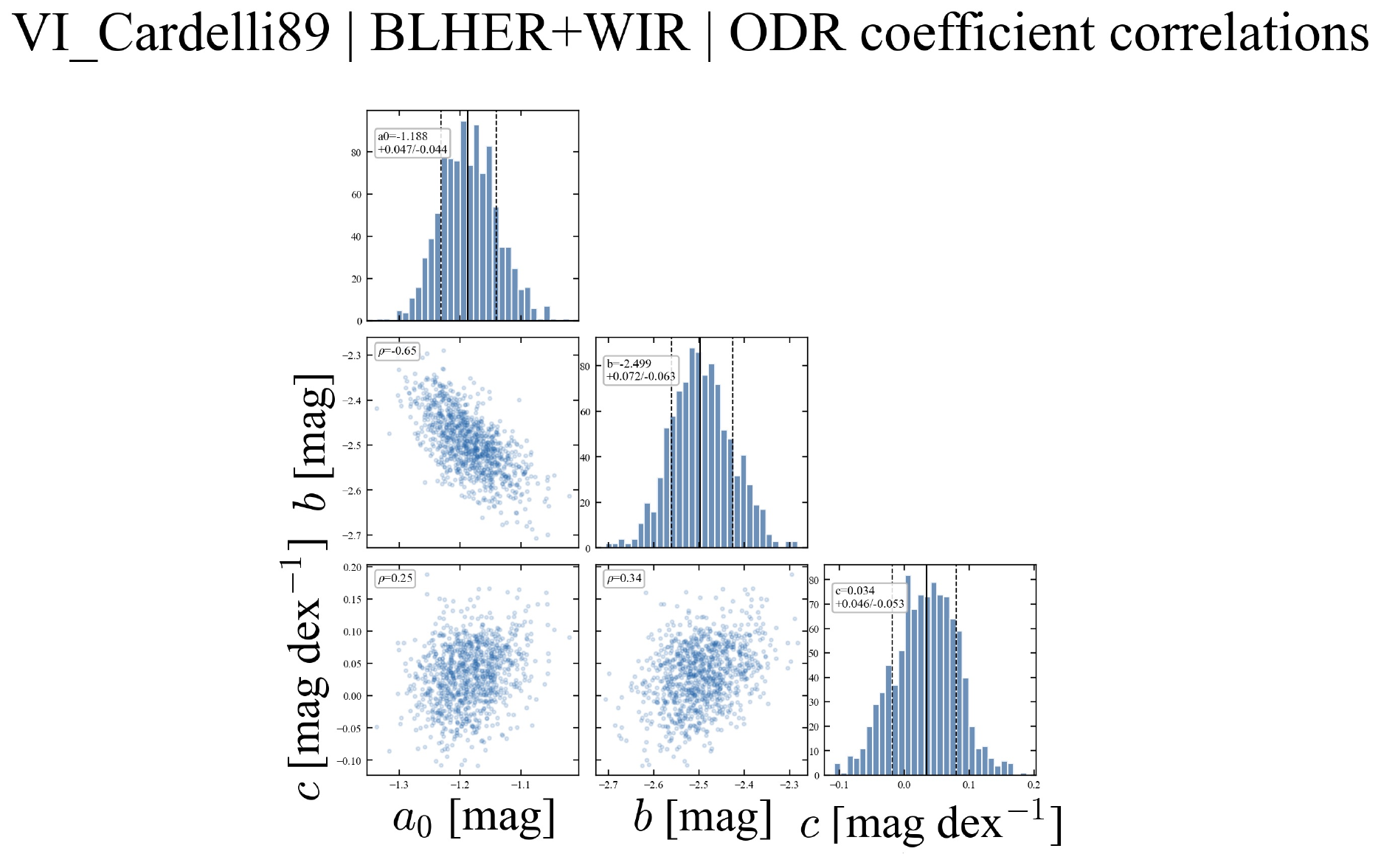}
    \caption{Same as Fig. \ref{cov 3} but for the PWZ relation on the VI bands based on the reddening law from \citet[][]{cardelli1989relationship}.} 
        \end{figure}

\begin{figure}
\centering
    \includegraphics[
    width=0.67\textwidth,
    trim=5.5cm 0.08cm 0cm 2cm,
    clip]{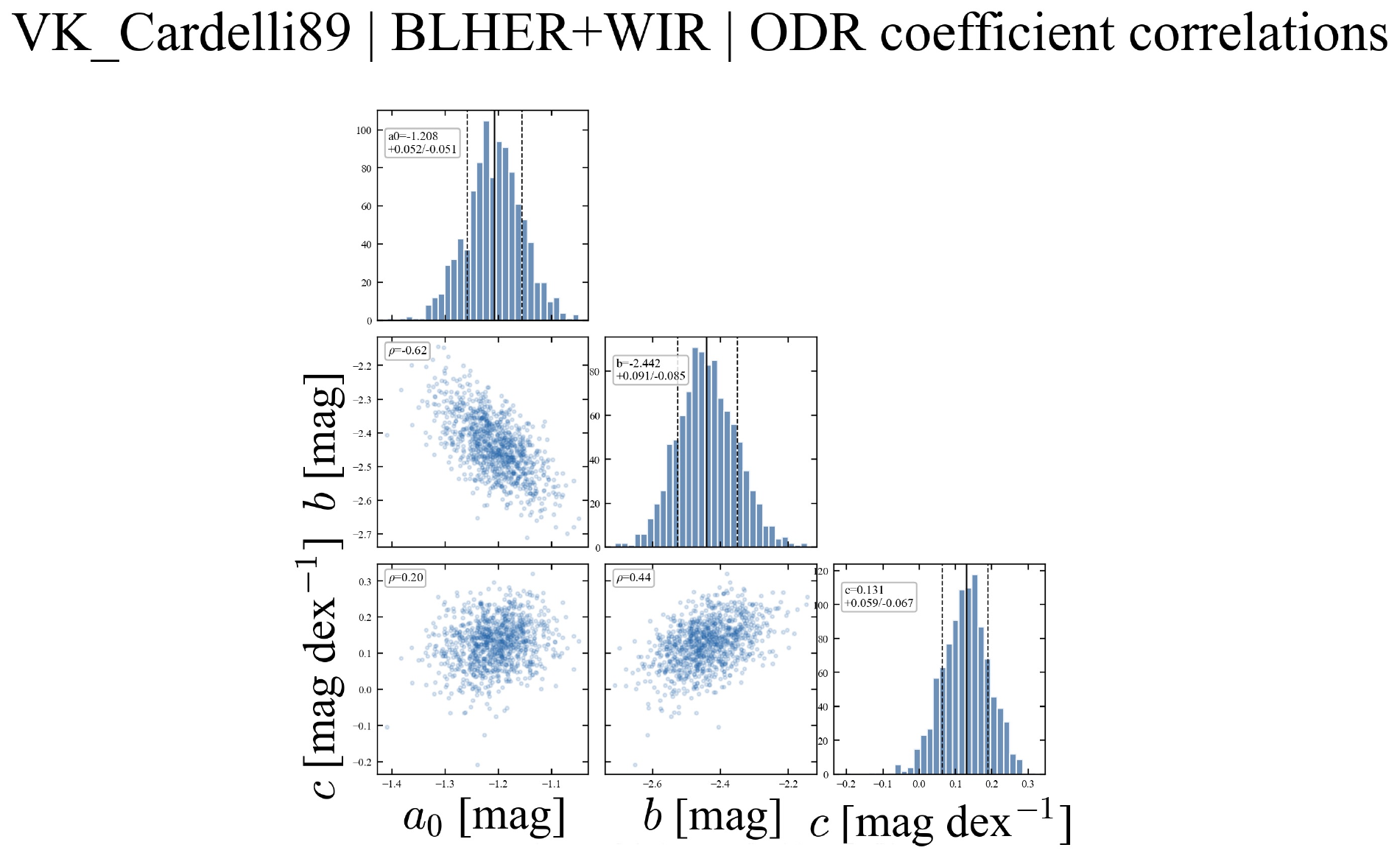}
    \caption{Same as Fig. \ref{cov 3} but for the PWZ relation on the VK bands based on the reddening law from \citet[][]{cardelli1989relationship}.} 
        \end{figure}

\begin{figure}
\centering
    \includegraphics[
    width=0.67\textwidth,
    trim=5.5cm 0.08cm 0cm 2cm,
    clip]{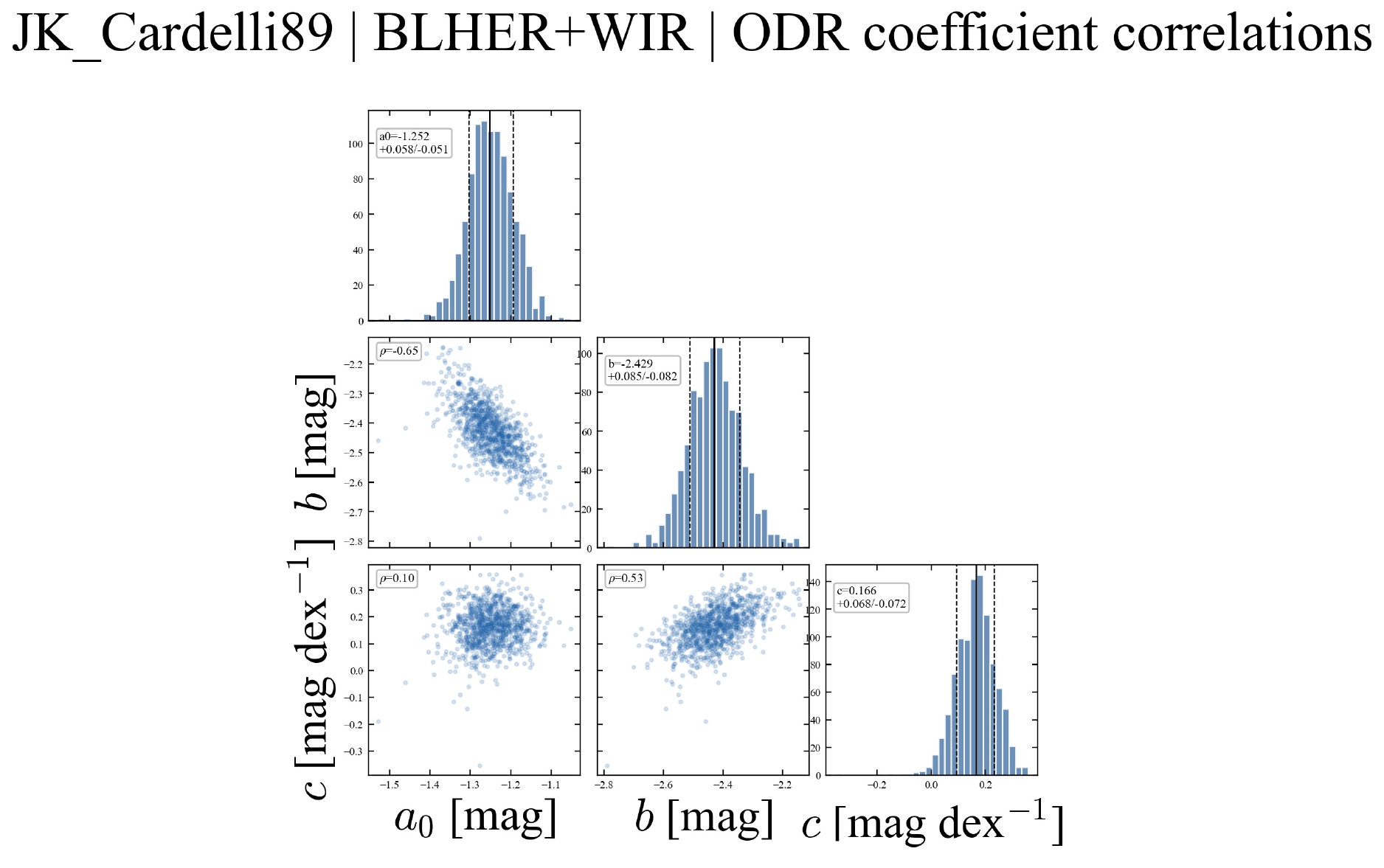}
    \caption{Same as Fig. \ref{cov 3} but for the PWZ relation on the JK bands based on the reddening law from \citet[][]{cardelli1989relationship}.} 
        \end{figure}






\FloatBarrier 
\clearpage

\end{appendix}
\end{document}